\documentclass[aps,prd,preprint,superscriptaddress,tightenlines,nofootinbib]{revtex4}

\usepackage{graphicx}
\usepackage{dcolumn}
\usepackage{bm}

\begin{document}

\preprint{CLEO CONF 06-9}   
\title{$\chi_{cJ}$ Decays to $h^+h^-h^0$\hspace*{2mm}}
\thanks{Submitted to the 33$^{\rm rd}$ International Conference on High Energy
Physics, July 26 - August 2, 2006, Moscow}


\author{S.~B.~Athar}
\author{R.~Patel}
\author{V.~Potlia}
\author{J.~Yelton}
\affiliation{University of Florida, Gainesville, Florida 32611}
\author{P.~Rubin}
\affiliation{George Mason University, Fairfax, Virginia 22030}
\author{C.~Cawlfield}
\author{B.~I.~Eisenstein}
\author{I.~Karliner}
\author{D.~Kim}
\author{N.~Lowrey}
\author{P.~Naik}
\author{C.~Sedlack}
\author{M.~Selen}
\author{E.~J.~White}
\author{J.~Wiss}
\affiliation{University of Illinois, Urbana-Champaign, Illinois 61801}
% ...
\collaboration{CLEO Collaboration} 
\noaffiliation

\date{July 24, 2006}

\begin{abstract} 
Using a sample of $3 \times 10^6$ $\psi(2S)$ decays recorded by the CLEO detector,
we study three body decays of the $\chi_{c0}$, $\chi_{c1}$, and
$\chi_{c2}$ produced in radiative decays of the $\psi(2S)$.  
We consider the decay modes $\pi^+\pi^-\eta$, $K^+K^-\eta$, $p{\bar p}\eta$,
$\pi^+\pi^-\eta^\prime$, $K^+K^-\pi^0$, $p{\bar p}\pi^0$, $\pi^+K^-K^0_{\rm S}$, and
$K^+{\bar p}\Lambda$ measuring branching fractions or placing upper limits.
For $\chi_{c1} \to \pi^+\pi^-\eta$,  $K^+K^-\pi^0$,
and $\pi^+K^-K^0_{\rm S}$ our observed samples are large enough to study the
substructure in a Dalitz plot analysis.
The results presented in this document are preliminary.
\end{abstract}

\maketitle


	Decays of the $\chi_{cJ}$ states are not as well studied both
experimentally and theoretically as those of other charmonium states.  
It is possible that the color-octet mechanism,
$c \bar c g \to 2(q \bar q)$ could have large effects on the observed
decay pattern of the $\chi_{cJ}$ states~\cite{quarkoniumreview}.
Thus any knowledge of any hadronic decay channels for these
state is valuable.

	CLEO has gathered a large sample of $e^+e^- \to \psi(2S)$
which leads to copious production of the $\chi_{cJ}$ states in
radiative decays of the $\psi(2S)$.  This contribution describes
our study of selected three body hadronic decay modes of the
$\chi_{cJ}$ to two charged and one neutral hadron.
This is not an exhaustive study of $\chi_{cJ}$ hadronic decays;
we do not even comprehensively cover all possible $h^+h^-h^0$ decays,
but simply take a first look
at the rich structure of $\chi_{cJ}$ decays in our initial
$\psi(2S)$ data sets.
With the CLEO III detector configuration~\cite{CLEOIII}, 
we have observed an integrated luminosity of 
2.57~pb$^{-1}$ and the number of $\psi(2S)$ events
is $1.56 \times 10^6$.
With the CLEO-c detector configuration~\cite{CLEOc} we have observed
2.89~pb$^{-1}$, and the number
of events is $1.52 \times 10^6$. Note that the apparent mis-match of luminosities 
and event totals is due to different beam energy spreads for the two data sets.  

	Our basic technique is an exclusive whole event analysis.  A photon
candidate is combined with three hadrons and the 4-momentum sum constrained
to the known beam energy and small beam momentum caused by the
beam crossing angle taking into account the measured errors on
the reconstructed charged tracks, neutral hadron, and transition
photon.  We cut on the $\chi^2$ of this fit, which has four degrees of freedom,
as it strongly 
discriminates between background and signal.  Efficiencies
and backgrounds are studied in a GEANT-based simulation~\cite{cleog} of
the detector response to underlying $e^+e^- \to \psi(2S)$ events.
Our simulated sample is roughly ten times our data sample.
The simulation is generated with a $ 1 + \lambda \cos^2\theta$
distribution in $\cos\theta$, where $\theta$ is the radiated photon angle 
relative to the positron beam axis. 
A E1 transition, as expected for $\psi(2S) \to \gamma\chi_{cJ}$,
implies $\lambda=1,-1/3,+1/13$ for $J=0,1,2$ particles. 
The efficiencies we quote use this simulation.
The differences of efficiencies 
due to various $\theta$ distributions are negligible
as we accept transition photons down to our detection limit.

A kinematic fit is made to each event.
For most modes we select events with an event 4-momentum fit 
$\chi^2$ less than 25, but
background from $\psi(2S)\to J/\psi\pi^0\pi^0$ followed by charged two
body decays of the $J/\psi$ with one of the $\pi^0$
decay photons lost fakes $\psi(2S)\to\gamma\chi_{cJ}\to \gamma p{\bar p}\pi^0$.
For this mode the cut on $\chi^2$ is tightened to 12.  The
efficiency of this cut is $\approx$ 95\% for all modes except
$p{\bar p}\pi^0$ where it is $\approx$ 80\%.  Little background
survives.

	Photon candidates are energy depositions in our crystal calorimeter
that have a transverse shape consistent with that expected for an electromagnetic
shower without a charged track pointing toward it.  They have an energy of at
least 30 MeV.  Transition photon candidates are vetoed if they form a
$\pi^0$ or $\eta$ candidate when paired with a second photon candidate.
Photon candidates that are formed into neutral particles further must
have an energy of more than 50 MeV if they are not in the central
barrel of our calorimeter.  $\pi^0 \to \gamma\gamma$ and $\eta \to \gamma\gamma$
candidates are formed from two photon candidates that are kinematically fit to
the known resonance masses
using the event vertex position, determined using charged tracks constrained to
the beam spot.  We select those giving a $\chi^2$ from the kinematic mass
fit with one degree of freedom of less than 10.

We also select the $\eta \to \pi^+\pi^-\pi^0$ mode combining
two charged pions with a $\pi^0$ candidate increasing the efficiency
for $\eta$ reconstruction by about 25\%.  The same sort of kinematic mass fit
as used for $\pi^0$'s and $\eta \to \gamma\gamma$
is applied to this mode, and again we select those giving a $\chi^2$
of less than 10.
Similarly we combine the mass-constrained $\eta$'s together with two charged pions to 
make $\eta^{\prime}$ candidates, mass constrain them, and select those with $\chi^2 < 10$. 
In addition, we include the decay mode $\eta^{\prime} \to \gamma \rho$. 
Here the background is potentially high because of the large 
number of noise photons, so we require 
$E_{\rm photon} > 200$~MeV. In addition, we require the $\pi^+\pi^-$ mass
to be within 100~MeV/c$^2$ of the $\rho$ mass.  

Charged tracks have standard requirements that they
be of good fit quality.
Those coming from the origin must have an impact parameter
with respect to the beam spot less than the
maximum of $(5.0-(3.8\cdot p))$~mm or 1.2~mm, where $p$ is the measured
track momentum in GeV/c.
$K^0_S \to \pi^+\pi^-$ and $\Lambda \to p\pi^-$ candidates
are formed from good quality tracks that are constrained to come
from a common vertex.
The $K^0_S$ flight path is required to be greater than 5~mm and the 
$\Lambda$ flight path greater than 3~mm.  The mass cut around the $K^0_S$ mass is 
$\pm 10$~MeV/c$^2$, and around the $\Lambda$ mass $\pm 5$~MeV/c$^2$.
Events with only the exact number of selected tracks
are accepted.
This selection is very efficient, $>$99.9\%, for events
passing all other requirements.

	Pions are required to have specific ionization,
$dE/dx$, in our main drift chamber within four standard
deviations of the expected value for
a real pion at the measured momentum. 
For kaons and protons, a combined $dE/dx$ and RICH 
likelihood is formed and kaons are simply required
to be more kaon-like than pion or proton-like, 
and similarly for protons.
Cross feed between hadron species is negligible after all
other requirements. 

In modes comprising only two charged particles, 
there are some extra cuts to eliminate QED background which
produce charged leptons in the final state.
Events are rejected if the sum over all the charged
tracks produces a penetration into the muon system of more
than five interaction lengths.
Events are rejected if any track has $0.92<E/p<1.05$ and it
has a $dE/dx$ consistent with an electron.
This latter cut is not used for $p\bar{p}$ 
modes because anti-protons tend to deposit all their
energy in the calorimeter.  These cuts are essentially 100\% efficient
for the signal,
and do eliminate the small QED background.

The efficiencies averaged over the CLEO~III and CLEO-c
data sets for each mode including the branching fractions
$\eta \to \gamma\gamma$, $\eta \to \pi^+\pi^-\pi^0$, and $\eta^\prime \to \eta\pi^+\pi^-$
are given in Tables~\ref{tab:BRfits0}-\ref{tab:BRfits2} for
$\chi_{c0}$,
$\chi_{c1}$, and
$\chi_{c2}$ respectively.
\begin{figure}
\includegraphics*[width=5.5in]{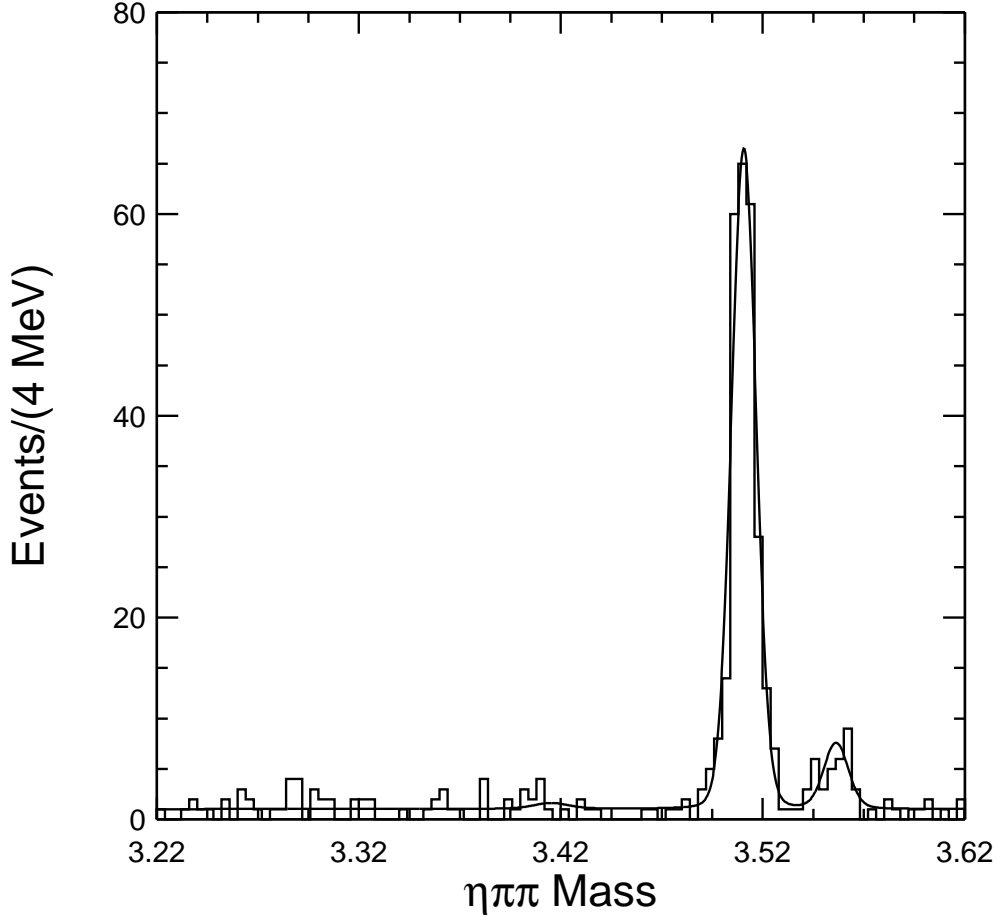}
\caption{Mass distribution for candidate $\chi_{cJ} \to \pi^+\pi^-\eta$.
         The displayed fit is described in the text.}
\label{fig:pipieta}
\end{figure}
\begin{figure}
\includegraphics*[width=5.5in]{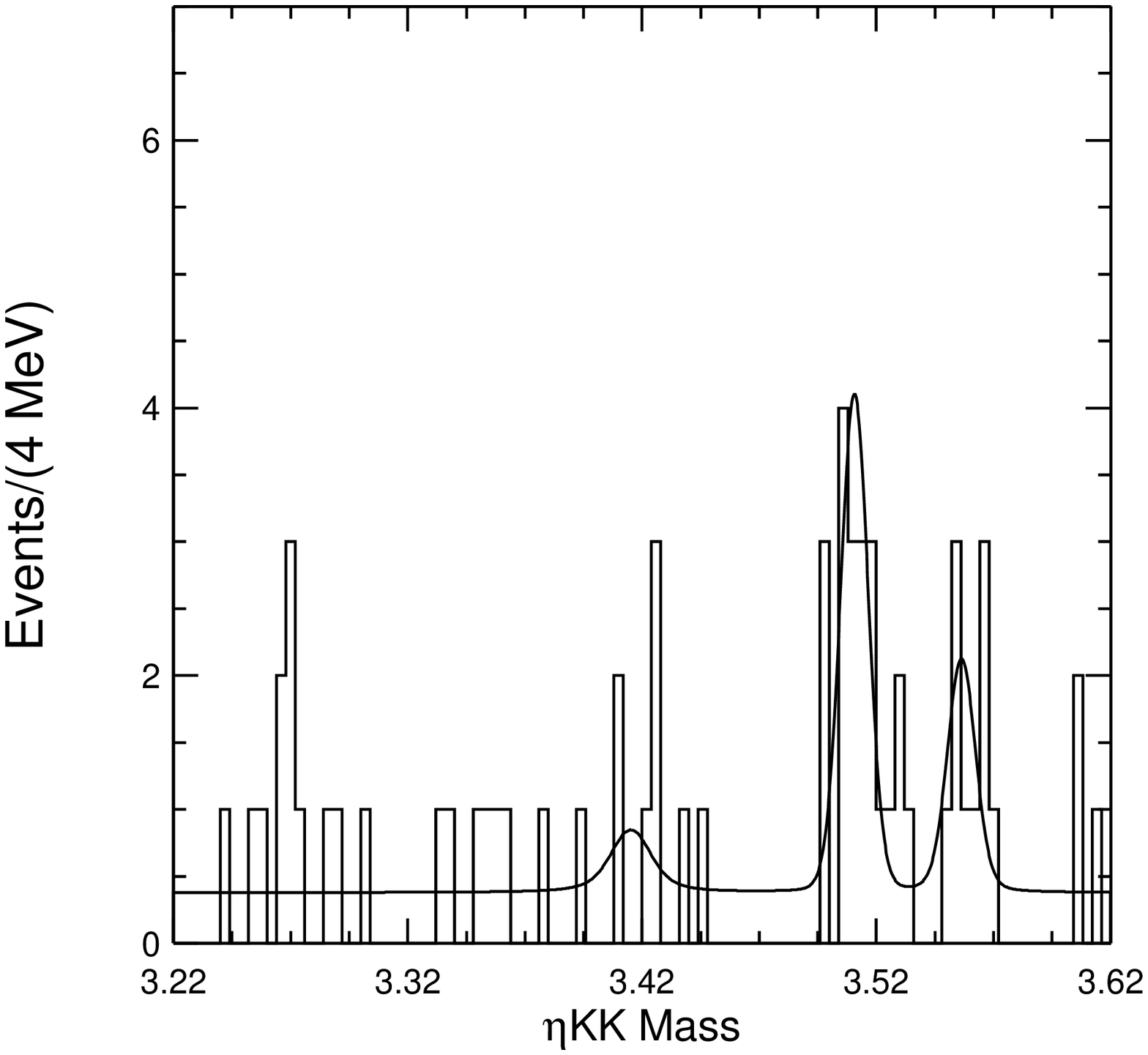}
\caption{Mass distribution for candidate $\chi_{cJ} \to K^+K^-\eta$.
         The displayed fit is described in the text.}
\label{fig:KKeta}
\end{figure}
\begin{figure}
\includegraphics*[width=5.5in]{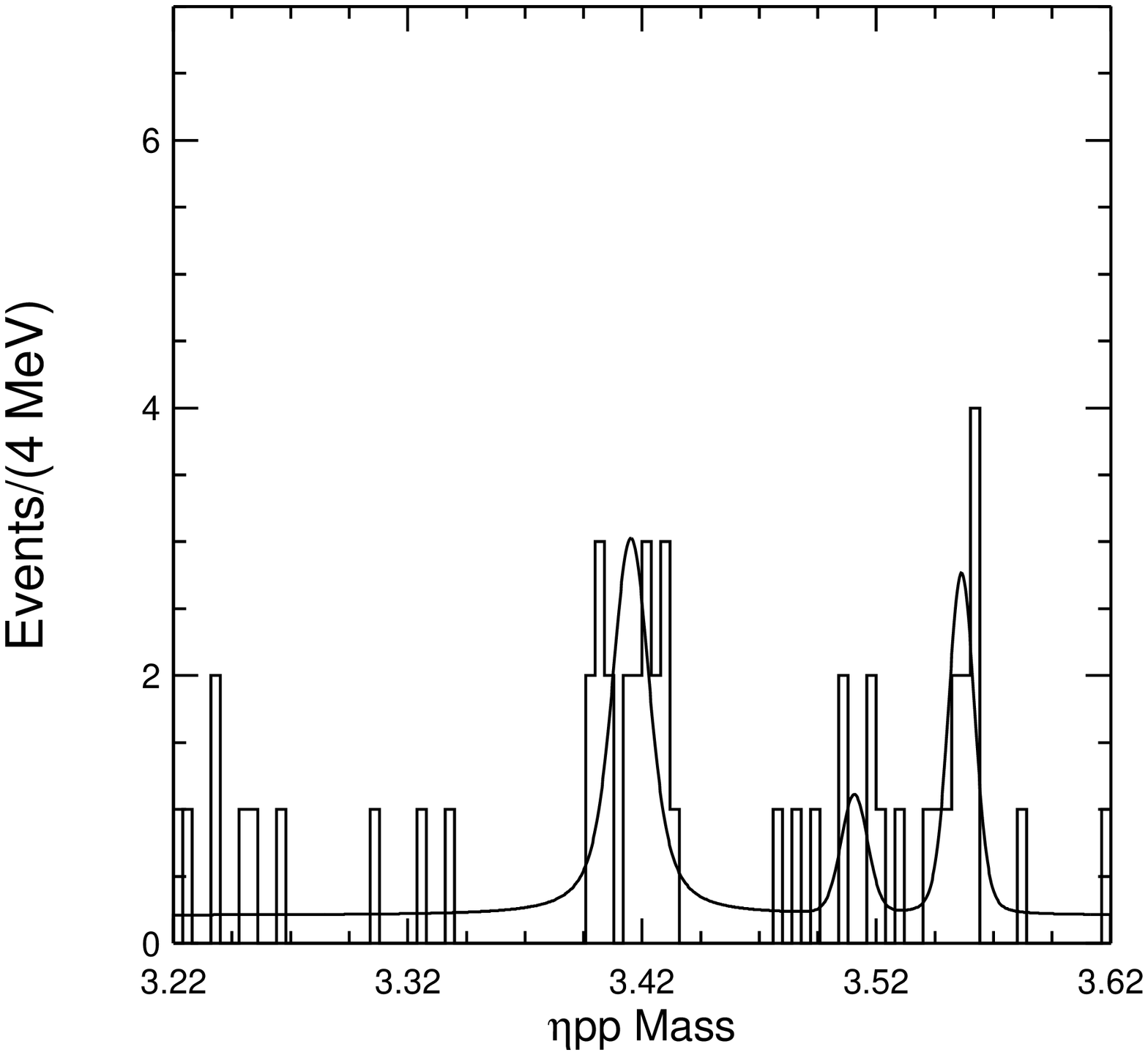}
\caption{Mass distribution for candidate $\chi_{cJ} \to p{\bar p}\eta$.
         The displayed fit is described in the text.}
\label{fig:ppbeta}
\end{figure}
\begin{figure}
\includegraphics*[width=5.5in]{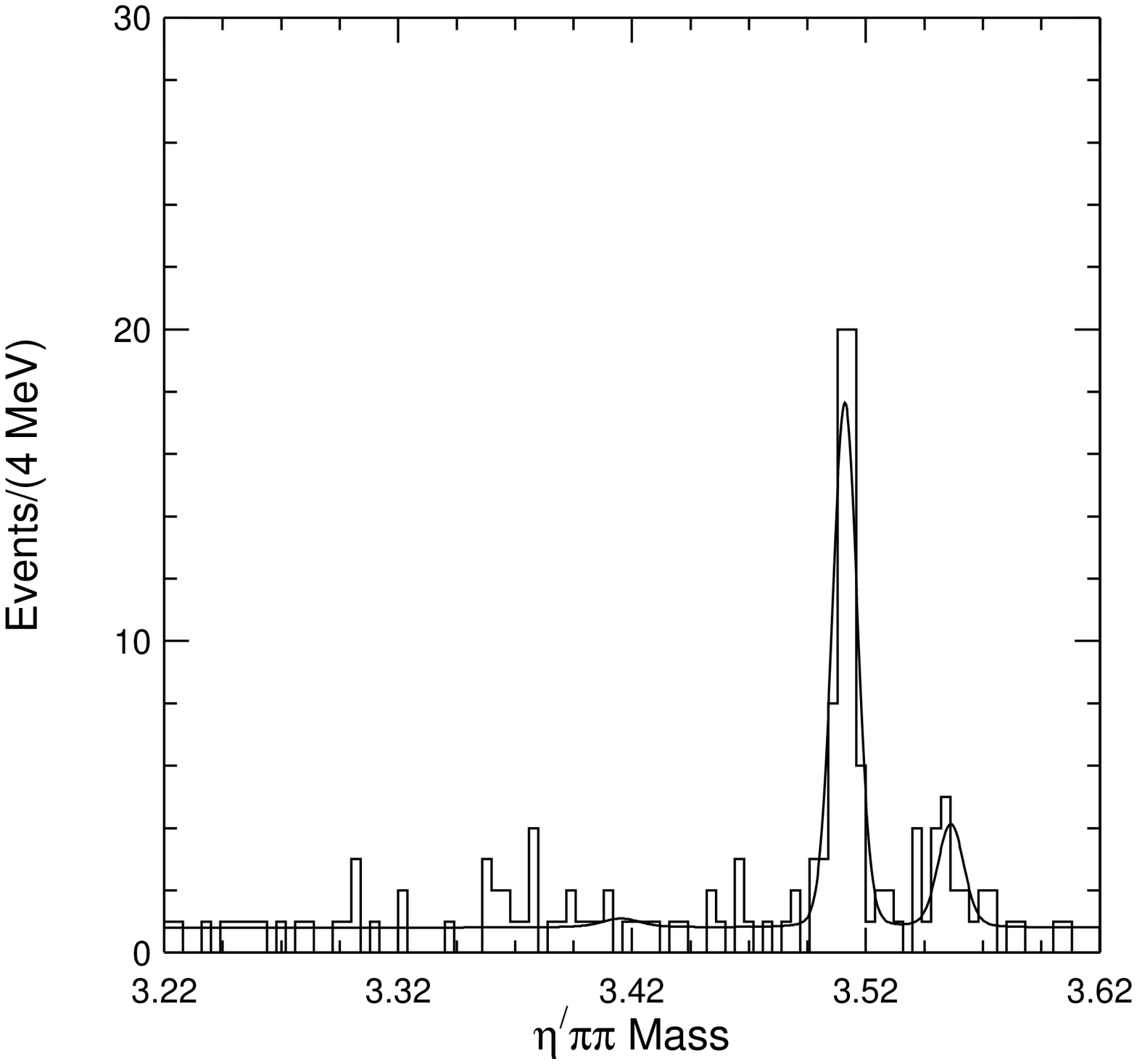}
\caption{Mass distribution for candidate $\chi_{cJ} \to \pi^+\pi^-\eta^\prime$.
         The displayed fit is described in the text.}
\label{fig:pipietaP}
\end{figure}
\begin{figure}
\includegraphics*[width=5.5in]{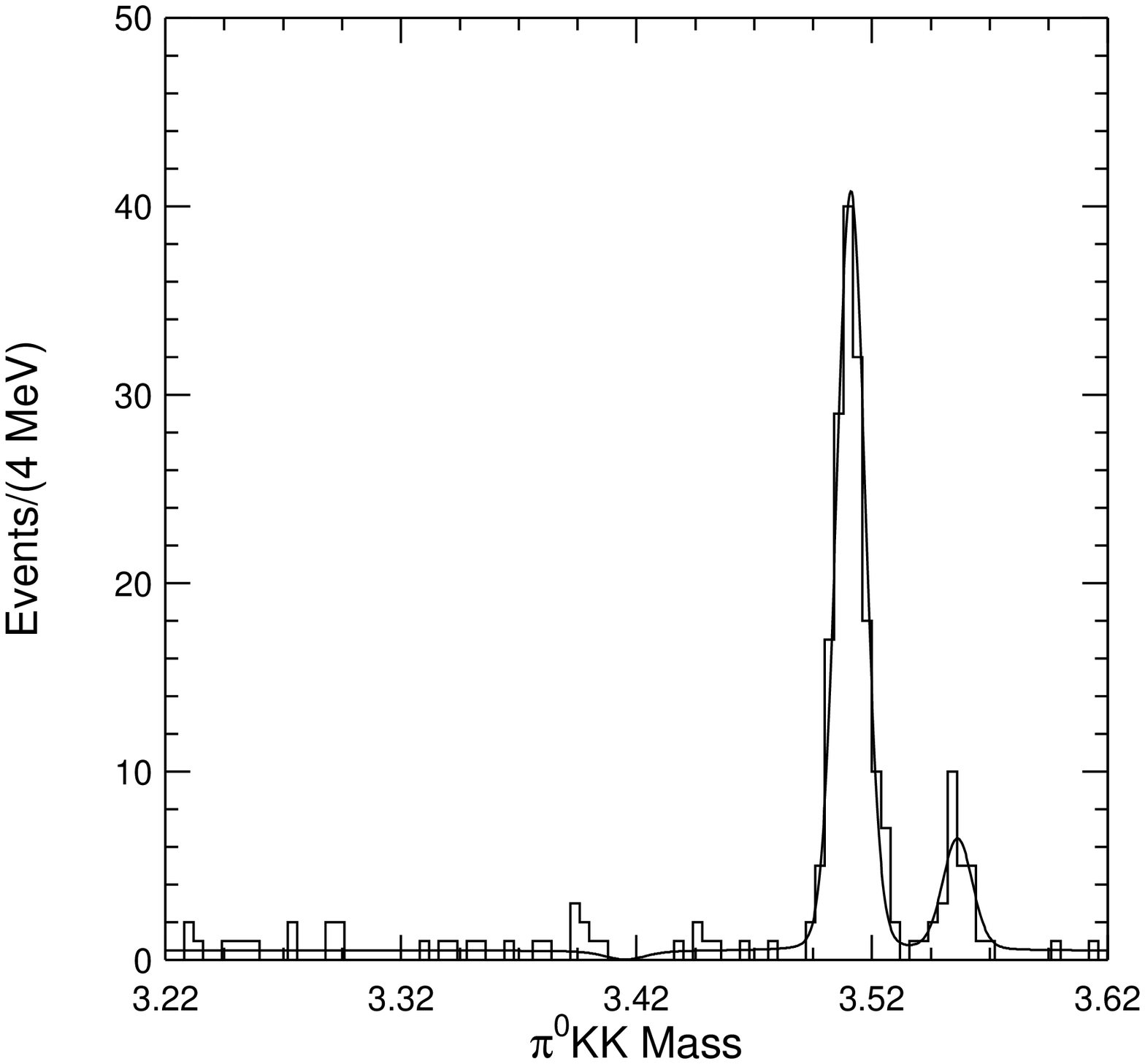}
\caption{Mass distribution for candidate $\chi_{cJ} \to K^+K^-\pi^0$.
         The displayed fit is described in the text.}
\label{fig:KKpi0}
\end{figure}
\begin{figure}
\includegraphics*[width=5.5in]{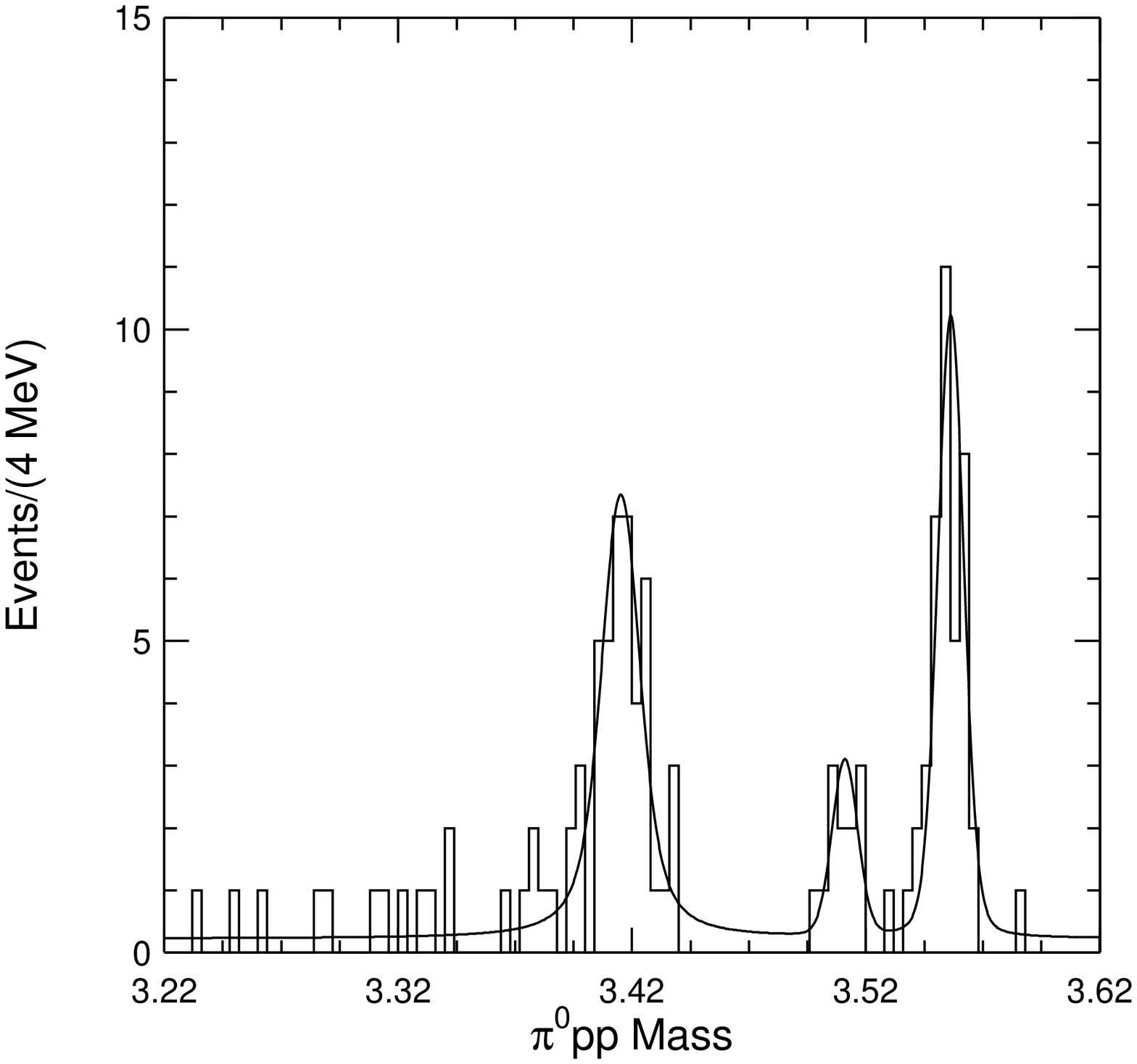}
\caption{Mass distribution for candidate $\chi_{cJ} \to p {\bar p} \pi^0$.
         The displayed fit is described in the text.}
\label{fig:ppbpi0}
\end{figure}
\begin{figure}
\includegraphics*[width=5.5in]{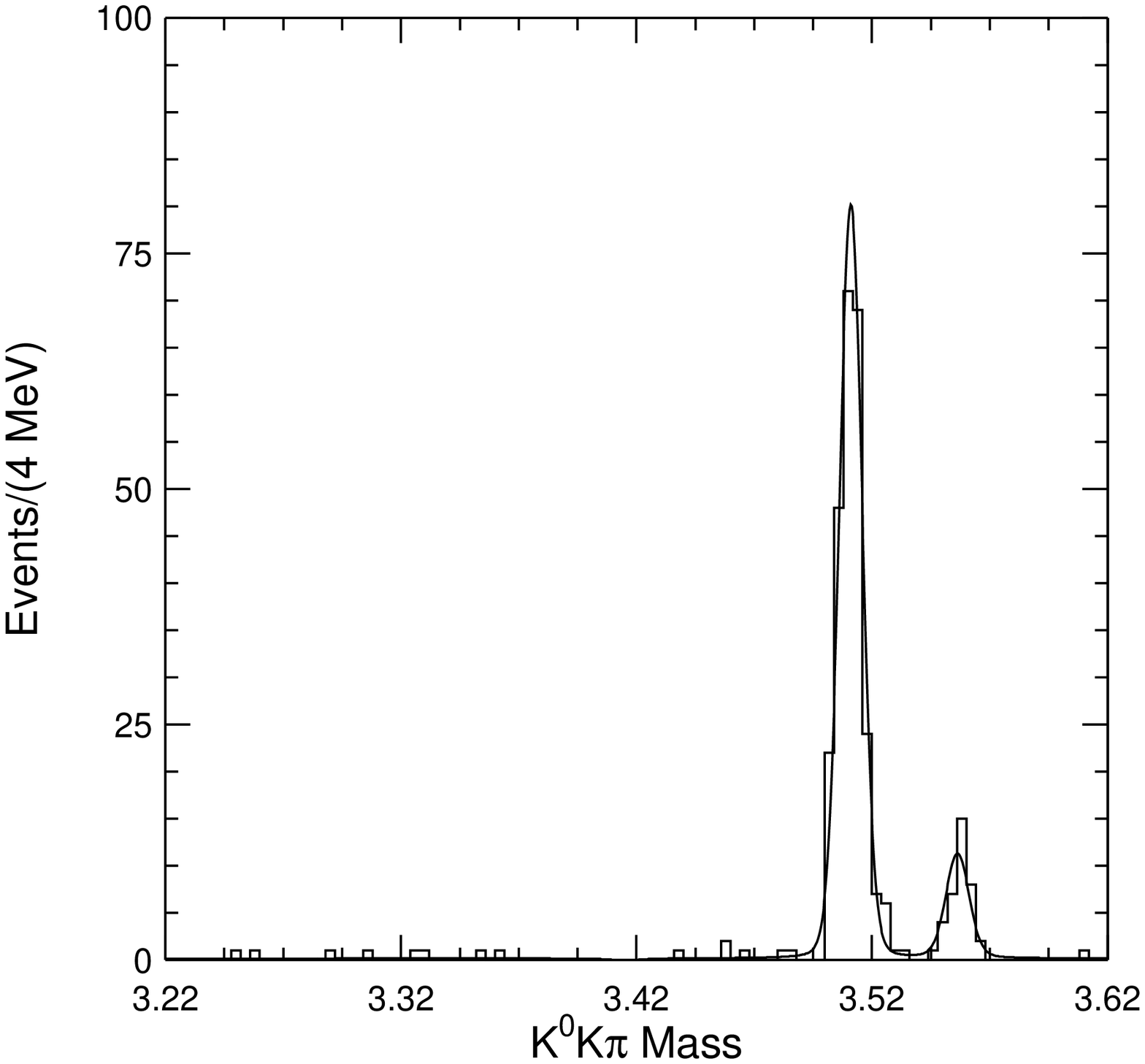}
\caption{Mass distribution for candidate $\chi_{cJ} \to K^0_{\rm S}K^-\pi^+$.
         The displayed fit is described in the text.}
\label{fig:piKKs}
\end{figure}
\begin{figure}
\includegraphics*[width=5.5in]{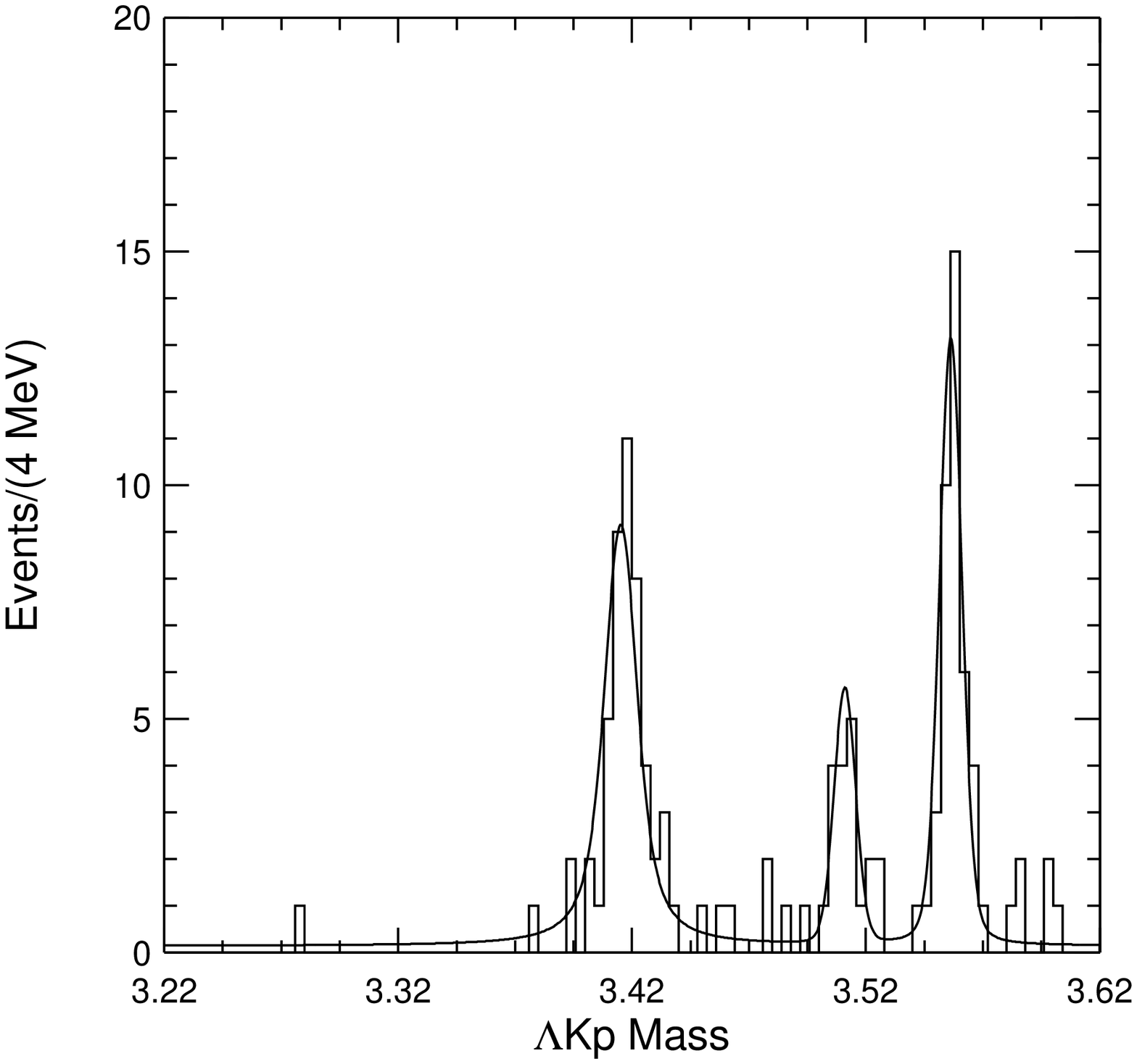}
\caption{Mass distribution for candidate $\chi_{cJ} \to K^+{\bar p}\Lambda$.
         The displayed fit is described in the text.}
\label{fig:Kpblam}
\end{figure}
Figures \ref{fig:pipieta}-\ref{fig:Kpblam} show the mass
distributions for the eight $\chi_{cJ}$ decay modes selected by
the analysis described above.  Signals are evident in all
three $\chi_{cJ}$ states, but not in all the modes.  Backgrounds
are small.  The mass distributions are fit to three signal shapes,
Breit-Wigners convolved with Gaussian detector resolutions, and a linear
background.  The $\chi_{cJ}$ masses and intrinsic 
widths are fixed at the values from the Particle
Data Group compilation~\cite{pdg}.  The detector resolution is taken
from the simulation discussed above.  The simulation
properly takes into account the amount of data in 
the two detector configurations, and the distribution
of different decay modes we have observed.  We approximate
the resolution with a single Gaussian distribution, and variations
are considered in the determination of systematic uncertainty.
The detector resolution dominates for the $\chi_{c1}$ and $\chi_{c2}$,
but is similar to the intrinsic width of the $\chi_{c0}$.
The fits are displayed in Figures \ref{fig:pipieta}-\ref{fig:Kpblam}
and summarized in Tables~\ref{tab:BRfits0}-\ref{tab:BRfits2}.
\begin{table}
\caption{Parameters used and results of fits to the $\chi_{cJ}$
         mass distributions of Figures~\ref{fig:pipieta}-\ref{fig:Kpblam} for the $\chi_{c0}$
         signal. The
         fit is described in the text and the yield error is only statistical.
         If no significant signal is observed the 90\% confidence level limit is also shown.}
\begin{tabular}{l|c|c|l}
Mode                     & Efficiency (\%) & Resolution (MeV) & Yield \\ \hline
$\pi^+\pi^-\eta$         & 17.6            & 6.23             & $4.0\pm4.1$ ($<10.6$)\\
$K^+K^-\eta$             & 13.6            & 6.10             & $3.1\pm2.7$ ($<9.3$) \\
$p{\bar p}\eta$          & 15.8            & 5.42             & $17.7^{+5.2}_{-4.8}$ \\
$\pi^+\pi^-\eta^\prime$  &  7.8            & 4.38             & $1.8\pm3.8$ ($<8.5$) \\
$K^+K^-\pi^0$            & 26.4            & 6.47             & $-3.3\pm2.7$ ($<4.7$)\\
$p{\bar p}\pi^0$         & 27.1            & 5.80             & $46.2^{+8.0}_{-7.2}$ \\
$\pi^+K^-K^0_{\rm S}$    & 19.0            & 4.75             & $0.0\pm1.0$ ($<2.7$)             \\
$K^+{\bar p}\Lambda$     & 16.7            & 4.38             & $51.2^{+8.1}_{-7.4}$ \\
\end{tabular}
\label{tab:BRfits0} 
\end{table}
\begin{table}
\caption{Parameters used and results of fits to the $\chi_{cJ}$
         mass distributions of Figures~\ref{fig:pipieta}-\ref{fig:Kpblam} for the $\chi_{c1}$
         signal. The
         fit is described in the text and the yield error is only statistical.
         If no significant signal is observed the 90\% confidence level limit is also shown.}
\begin{tabular}{l|c|c|l}
Mode                     & Efficiency (\%) & Resolution (MeV) & Yield \\ \hline
$\pi^+\pi^-\eta$         & 18.1            & 5.85             & $255^{+17}_{-16}$    \\
$K^+K^-\eta$             & 14.2            & 5.66             & $14.1^{+4.6}_{-3.9}$ \\
$p{\bar p}\eta$          & 17.6            & 5.41             & $3.2\pm2.3$ ($<7.6$) \\
$\pi^+\pi^-\eta^\prime$  &  8.3            & 4.37             & $57.6^{+8.4}_{-7.7}$ \\
$K^+K^-\pi^0$            & 28.2            & 6.79             & $157\pm13$           \\
$p{\bar p}\pi^0$         & 29.5            & 5.23             & $9.9^{+3.8}_{-3.2}$  \\
$\pi^+K^-K^0_{\rm S}$    & 19.8            & 4.37             & $249\pm16$           \\
$K^+{\bar p}\Lambda$     & 17.5            & 4.38             & $16.3^{+4.7}_{-4.0}$ \\
\end{tabular}
\label{tab:BRfits1}
\end{table}
\begin{table}
\caption{Parameters used and results of fits to the $\chi_{cJ}$
         mass distributions of Figures~\ref{fig:pipieta}-\ref{fig:Kpblam} for the $\chi_{c2}$
         signal. The
         fit is described in the text and the yield error is only statistical.
         If no significant signal is observed the 90\% confidence level limit is also shown.}
\begin{tabular}{l|c|c|l}
Mode                     & Efficiency (\%) & Resolution (MeV) & Yield \\ \hline
$\pi^+\pi^-\eta$         & 17.8            & 5.58             & $26.2^{+6.4}_{-5.7}$  \\
$K^+K^-\eta$             & 14.1            & 5.53             & $6.9\pm2.9$ ($<12.54$)\\
$p{\bar p}\eta$          & 16.9            & 5.08             & $9.5^{+3.8}_{-3.0}$   \\
$\pi^+\pi^-\eta^\prime$  &  8.2            & 4.32             & $12.4^{+4.8}_{-4.1}$  \\
$K^+K^-\pi^0$            & 27.5            & 6.85             & $24.8^{+5.8}_{-5.1}$  \\
$p{\bar p}\pi^0$         & 28.9            & 5.10             & $37.1^{+6.7}_{-6.1}$  \\
$\pi^+K^-K^0_{\rm S}$    & 17.2            & 4.45             & $36.8^{+6.6}_{-5.9}$  \\
$K^+{\bar p}\Lambda$     & 17.5            & 4.32             & $42.1^{+7.2}_{-6.5}$  \\
\end{tabular}
\label{tab:BRfits2}
\end{table}
Note that for the $\chi_{c0}$ in Table~\ref{tab:BRfits0} the five modes for which no
significant signal are found
are forbidden by parity conservation.

\begin{figure}
\begin{tabular}{ccc}
\includegraphics*[width=2.0in]{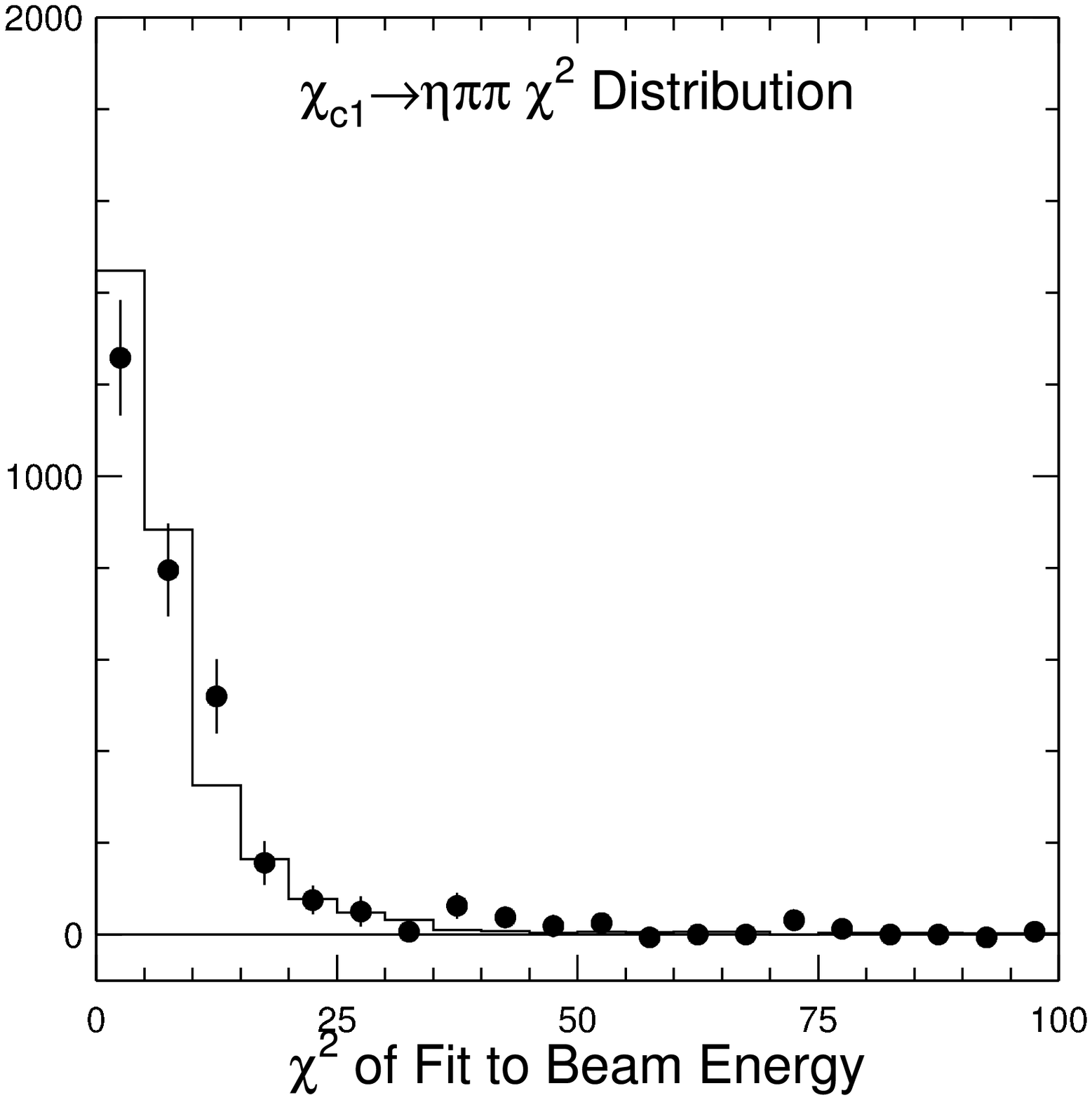} &
\includegraphics*[width=2.0in]{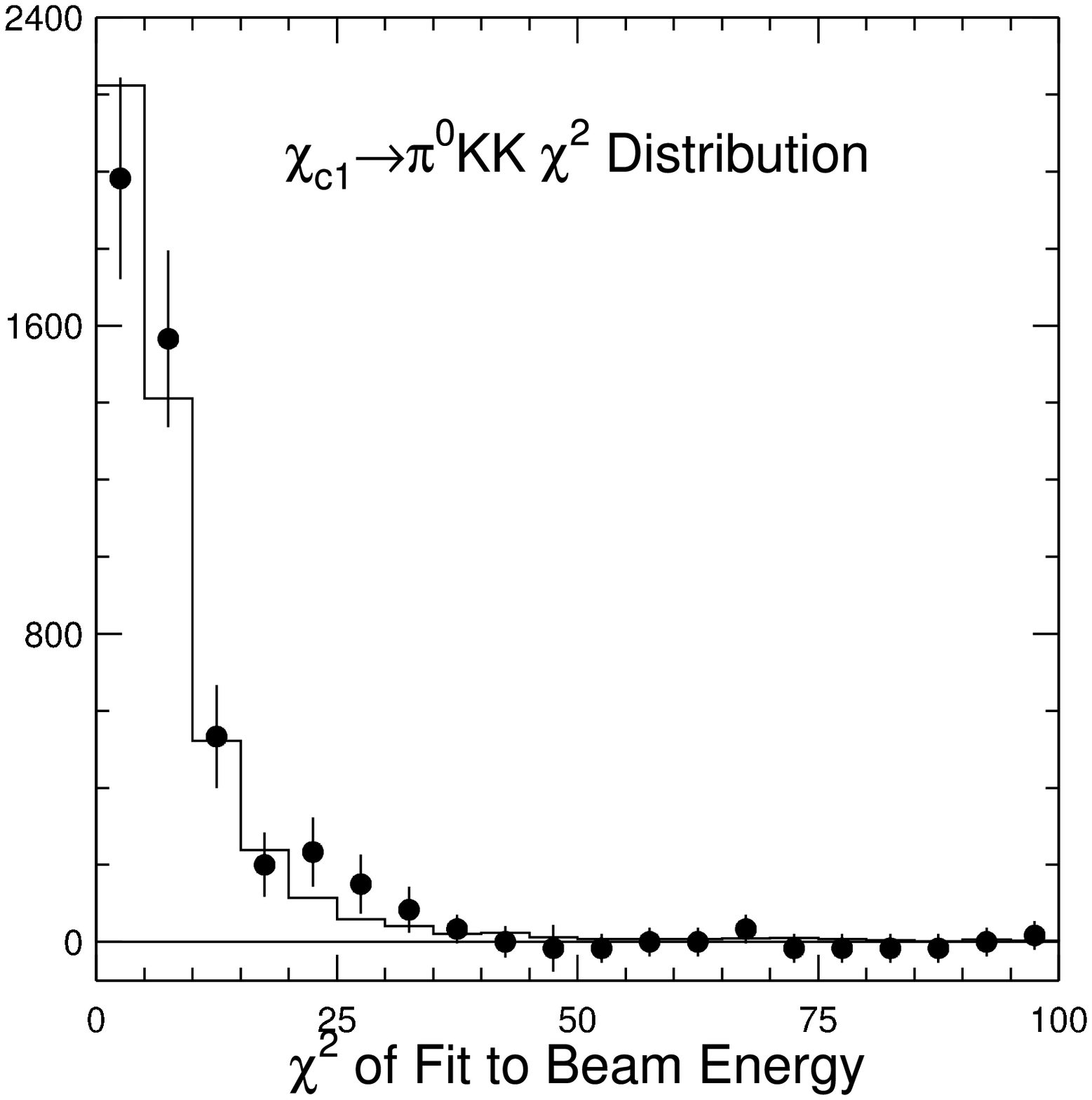} &
\includegraphics*[width=2.0in]{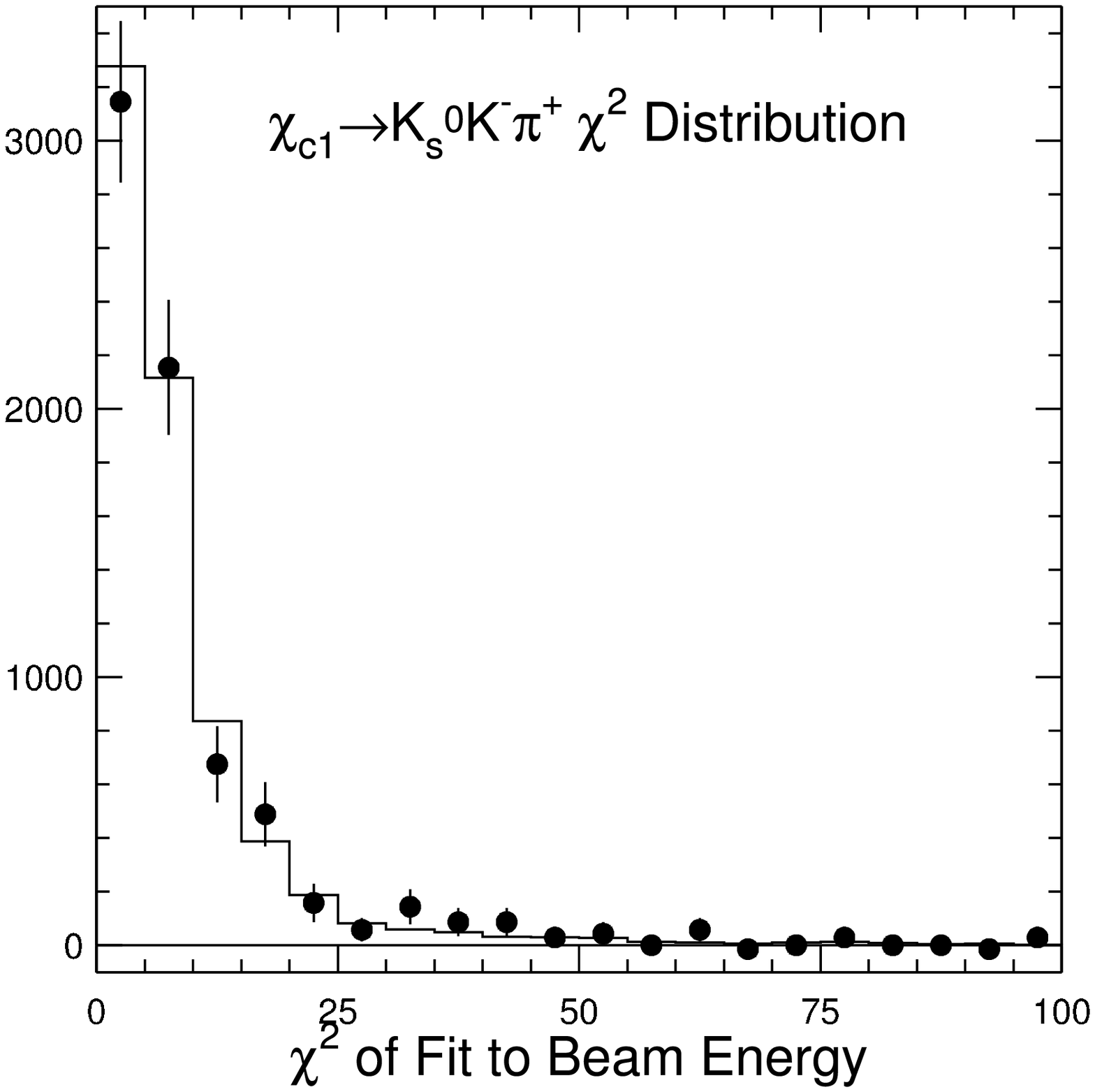}
\end{tabular}
\caption{Distribution of the $\chi^2$ of the beam energy constrained mass fit.
                        This is shown with a mass cut around the $\chi_{c1}$ signal region and after 
                        a data sideband subtraction.  The plot on the left is for the $\pi^+\pi^-\eta$
                        mode, center is the $K^+K^-\pi^0$ mode, and right is the $\pi^+K^-K^0_{\rm S}$
                        mode.  The data are shown by points and the simulation of signal events is
                        shown by the histogram.}
\label{fig:chi2}
\end{figure}

	We consider various sources of systematic uncertainties on the yields.
We varied the fitting procedure by allowing the $\chi_{cJ}$ masses and intrinsic widths
to float.  The fitted masses and widths agree with the fixed values from the particle
data group, and we take the maximum variation in the observed yields, 4\%, as a systematic
uncertainty from the fit procedure.  Allowing a curvature term to the background
has a negligible effect.
For modes with large yields we can break up the sample into CLEO~III and CLEO-c
data sets, and fit with resolutions and efficiencies appropriate for the individual
data sets.  We note that the separate data sets give consistent efficiency
corrected yields and the summed yield differs by 2\% from the standard procedure,
which is small
compared to the 8\% statistical uncertainty.  We take this as the systematic
uncertainty from our resolution model.  From studies of other processes
we assign a 0.7\% uncertainty for the efficiency of finding each charged
track, 2.0\% for the $\gamma\gamma$ resonances, 1.0\% for each extra photon,
1.3\% for the particle identification for each $K$ and $p$, 2.0\% for
secondary vertex finding, and 3.0\% from the statistical uncertainty
on the efficiency determined from the simulation.  We study the cut on the $\chi^2$
of event 4-momentum kinematic fit in the three large yield $\chi_{c1}$ signals
by removing the $\chi^2$ cut, selecting events around the $\chi_{c1}$
mass peak, subtracting a low mass side band, the only one available, and 
comparing the simulated $\chi^2$ distribution for signal events with the data
distribution.  This comparison is shown in Figure~\ref{fig:chi2}.  The agreement
between the data and simulation is good, and comparing the inefficiency introduced
by our cut on the 4-momentum kinematic fit $\chi^2$ between the data and the simulation
we assign a 3.5\% uncertainty on the efficiency due to uncertainty in modeling
this $\chi^2$ distribution.  
The simulation was generated assuming 3-body phase-space for the $\chi_{cJ}$ decay products. 
Deviations from this are to be expected.  Based on the results of the Dalitz plot analyses
discussed below we correct the efficiency in the $\chi_{c1} \to \pi^+ \pi^-\eta $, $\chi_{c1} \to K^+K^-\pi^0$, and
$\chi_{c1} \to K^-\pi^+K^0_S$ modes by a relative $+6$\% to account for the change
in the efficiency caused by the deviation from a uniform phase space distribution
of decay products to what we actually observe.  This correction only has
a noticeable impact on the $\chi_{c1} \to K^-\pi^+K^0_S$ mode.

To calculate $\chi_{cJ}$ branching fractions, we use previous CLEO measurements
for the $\psi(2S)\to\chi_{cJ}$
branching fractions of $9.3 \pm 0.14 \pm 0.61$\%, $9.07 \pm .11\pm.54$\% and $9.22 \pm .11\pm.54$\%
for $J$=0,1,2 respectively~\cite{chicbf}.  The uncertainties on these branching fractions
are included in the systematic uncertainty on the $\chi_{cJ}$ branching fractions we report.
Preliminary results for the three body branching fractions are shown in 
Table~\ref{tab:Branching_fractions}.  Where the yields do not show clear signals
\begin{table}
\caption{\label{tab:Branching_fractions}
Preliminary branching fractions in \%.  Uncertainties are statistical,
systematic due to detector effects plus analysis methods
and a separate systematic due to uncertainties in the $\psi(2S)$ branching
fractions.  Limits are at the 90\% confidence level.}
\small
\begin{tabular}{l|c|c|c}
\hline
Mode &$\chi_{c0}$ & $\chi_{c1}$ & $\chi_{c2}$ \\
\hline
$\pi^+\pi^-\eta $         & $<0.021$  
                          & $0.52\pm .03\pm .03\pm .03$ 
                          & $0.051\pm .011\pm .004\pm .003$ \\
$K^+K^-\eta$              & $<0.024$  
                          & $0.034\pm .010\pm .003\pm .002$ 
                          & $<0.033$ \\
$p\bar{p}\eta$            & $0.038\pm .010\pm .003\pm .02$  
                          & $<0.015$ 
                          & $ .019\pm .007\pm .002\pm .002$ \\
$\pi^+\pi^-\eta^{\prime}$ & $<0.038$  
                          & $0.24\pm .03\pm .02\pm .02$     
                          & $<0.053$     \\
$K^+K^-\pi^0$             & $<0.006$  
                          & $0.200\pm .015\pm .018\pm .014$ 
                          & $0.032\pm .007\pm .002\pm .002$ \\
$p\bar{p}\pi^0$           & $0.059\pm .010\pm .006\pm .004$ 
                          & $0.014 \pm 0.005 \pm 0.001 \pm 0.001$ 
                          & $0.045\pm .007\pm 0.004\pm .003$ \\
$\pi^+K^-{\bar K^0}$      & $<0.010$  
                          & $0.84\pm .05\pm .06\pm .05$ 
                          & $0.15\pm .02\pm .01\pm .01$ \\
$K^+ \bar{p}\Lambda$      & $0.114\pm .016\pm .009\pm .007$ 
                          & $0.034\pm .009\pm .003\pm .002$ 
                          & $0.088\pm .014\pm .07\pm .006$ \\ 
\end{tabular}
\end{table}
we calculate 90\% confidence level upper limits using the yield central values with the statistical errors from the yield fits
combined in quadrature with the systematic uncertainties on the efficiencies and other branching fractions.  We assume
the uncertainty is distributed as a Gaussian and the upper limit is the branching fraction value at which 90\% of the
integrated area of the Gaussian falls below.  We exclude the unphysical region, negative branching fractions, for this
upper limit calculation.  We note that the ratio of rates expected from isospin symmetry, 
as discussed in Appendix~\ref{sec:Clebsch-Gordan},
Equations~\ref{eqn:case_width} and~\ref{eqn:case_width_for_a0},
expected to be 4.0 is
consistent with our measurement:
\begin{equation} \label{eqn:ratio_of_rates}
  \frac{\Gamma(\chi_{c1}\to \pi^+K^-K^0)
       +\Gamma(\chi_{c1}\to \pi^-K^+\overline{K^0}) }
       {\Gamma(\chi_{c1}\to \pi^0K^+K^-)}
       = 4.2 \pm 0.7.
\end{equation}

We choose the three high-statistics signals 
$\chi_{c1} \to \pi^+ \pi^-\eta $, $\chi_{c1} \to K^+K^-\pi^0$, and
$\chi_{c1} \to K^-\pi^+K^0_S$ for Dalitz plot analysis 
to study resonance substructure.  For the Dalitz analysis only those
events within 10 MeV, roughly two standard deviations, of the signal peak
are accepted.  For $\chi_{c1} \to \pi^+ \pi^-\eta$ there are 228 events in
this region and the signal fit finds 224.2 signal events and 5.1 
combinatorial background. For $\chi_{c1} \to K^+K^-\pi^0 $ there are
137 events accepted with the fit finding 137.8 signal and 2.4 background
events, and for $\chi_{c1} \to K^-\pi^+K^0_S$, the numbers
are 234 events, of which 233.2 are signal and 0.8 are background. In all
cases the contribution from the tail of the $\chi_{c2}$ is less than 
one event.

An unbinned maximum likelihood fit is used along with other methods
as described in~\cite{cleodalitz} in order to perform the Dalitz plot analysis. 
We only summarize our methods here.
Efficiencies are determined with simulated event samples generated
uniformly in phase space,
and run through the analysis procedure described above.
The efficiency across the Dalitz plots is fit to a two dimensional
polynomial of third order in the Dalitz plot variables.  The fits are
of good quality and the efficiency is generally flat across the 
Dalitz plot.

When fitting the data Dalitz plot the small contributions from backgrounds
are neglected.  We use an isobar model to describe resonance contributions
to the Dalitz plots taking into account spin and width dependent effects. 
Narrow resonances are described with a Breit-Wigner amplitude with the resonance
parameters taken from previous experiments \cite{pdg}.  For the scalar resonances 
$a_0(980)$ and $f_0(980)$ we use a Flatt\'e parameterization.
We use the $a_0(980)$ line-shape from the
Crystal Barrel Collaboration~\cite{CBarrel_a0_980} and the details of
the $f_0(980)$ are unimportant as it is used only in systematic studies.
For low mass $\pi^+\pi^-$($\sigma$) and  $K\pi$($\kappa$) S-wave contributions
we choose a simple description, one which is adequate for our small sample~\cite{Oller_2005}.

We are examining the $e^+e^- \to \psi(2S) \to \gamma \chi_{c1}$ process.
In such a decay the $\chi_{c1}$ should be polarized.  In principal
a more complete analysis would take into account the angle of the photon
with respect to the $e^+e^-$ beams collision axis
and decompose the $\chi_{c1}$ decay into
its partial waves.  We barely have the statistics to do a reasonable
Dalitz analysis suggesting that a higher dimensional partial wave 
analysis would be hopeless.  See Appendix~\ref{sec:angular_distributions}
for a discussion and the formalism of how polarization would affect the Dalitz
analysis when the intermediate resonance is not spin zero even when
integrated over the random polarization direction of the $\chi_{c1}$.
Keeping this in mind, 
for this Dalitz plot analysis we decided to use angular distributions 
from~\cite{Filippini-Fontana-Rotondi}.
We have tested different angular distributions and 
include the variations as a systematic uncertainty. 

\begin{figure}
\begin{tabular}{cc}
\includegraphics*[width=3.0in]{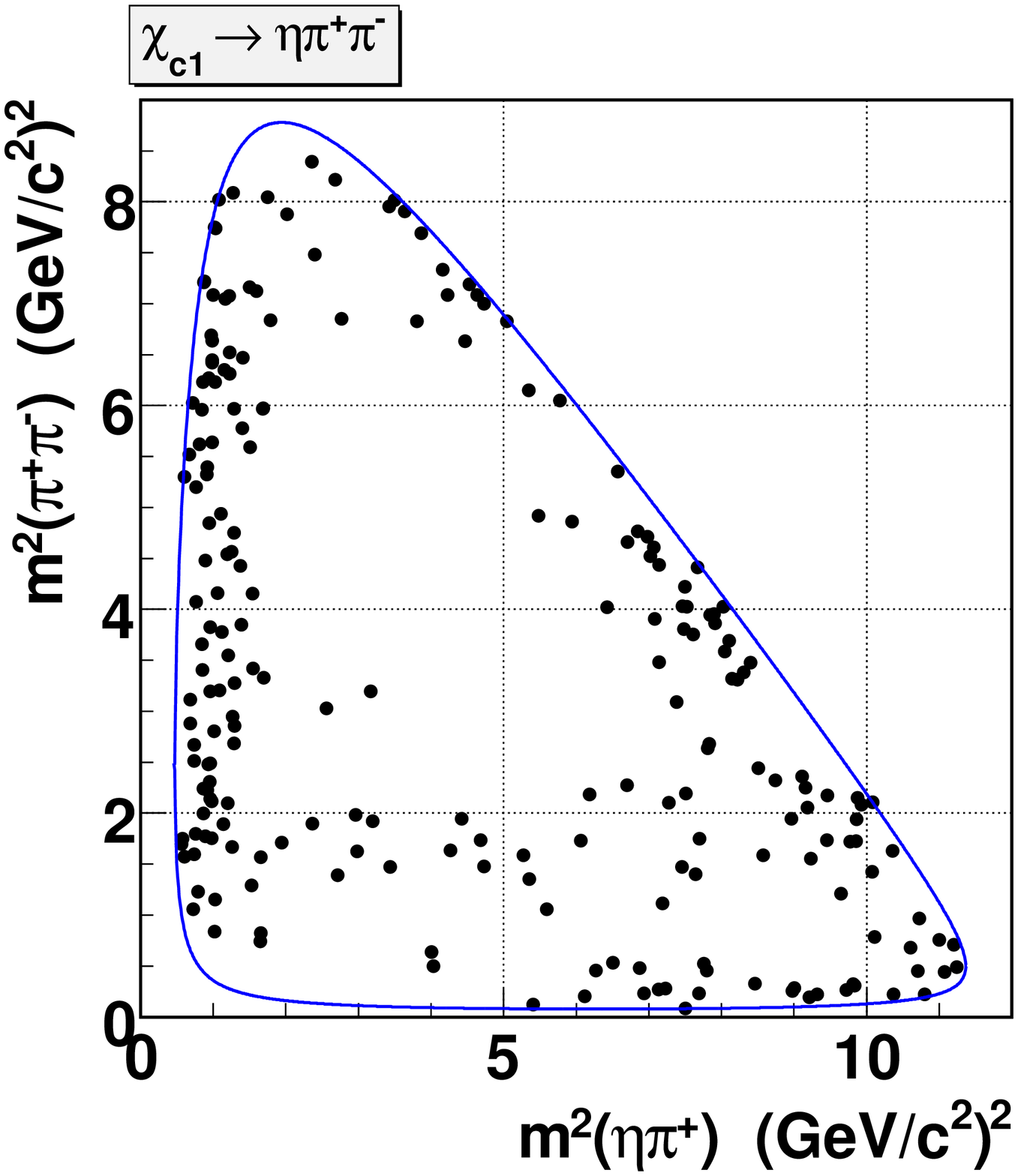} &
\includegraphics*[width=3.0in]{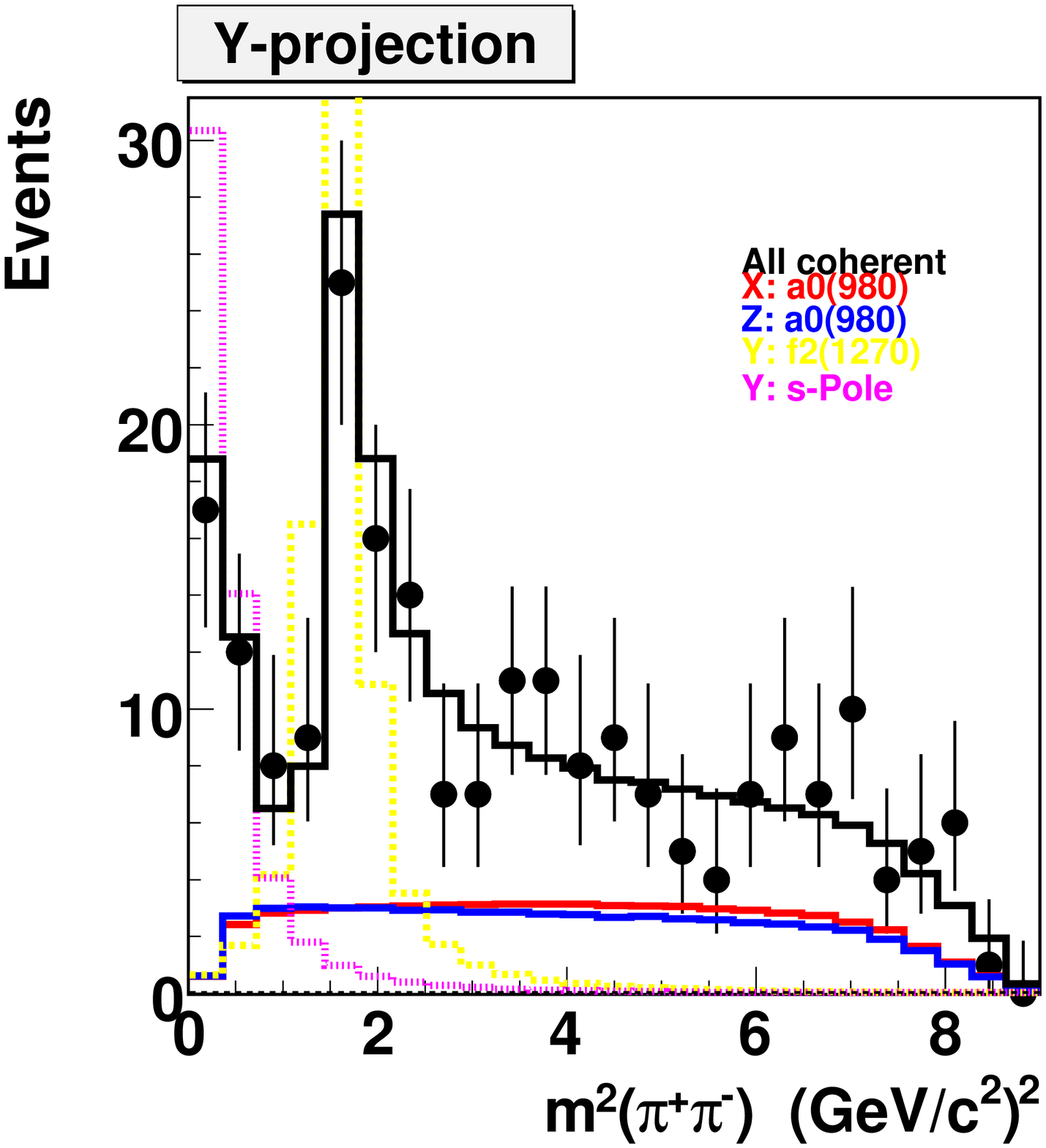} \\
\includegraphics*[width=3.0in]{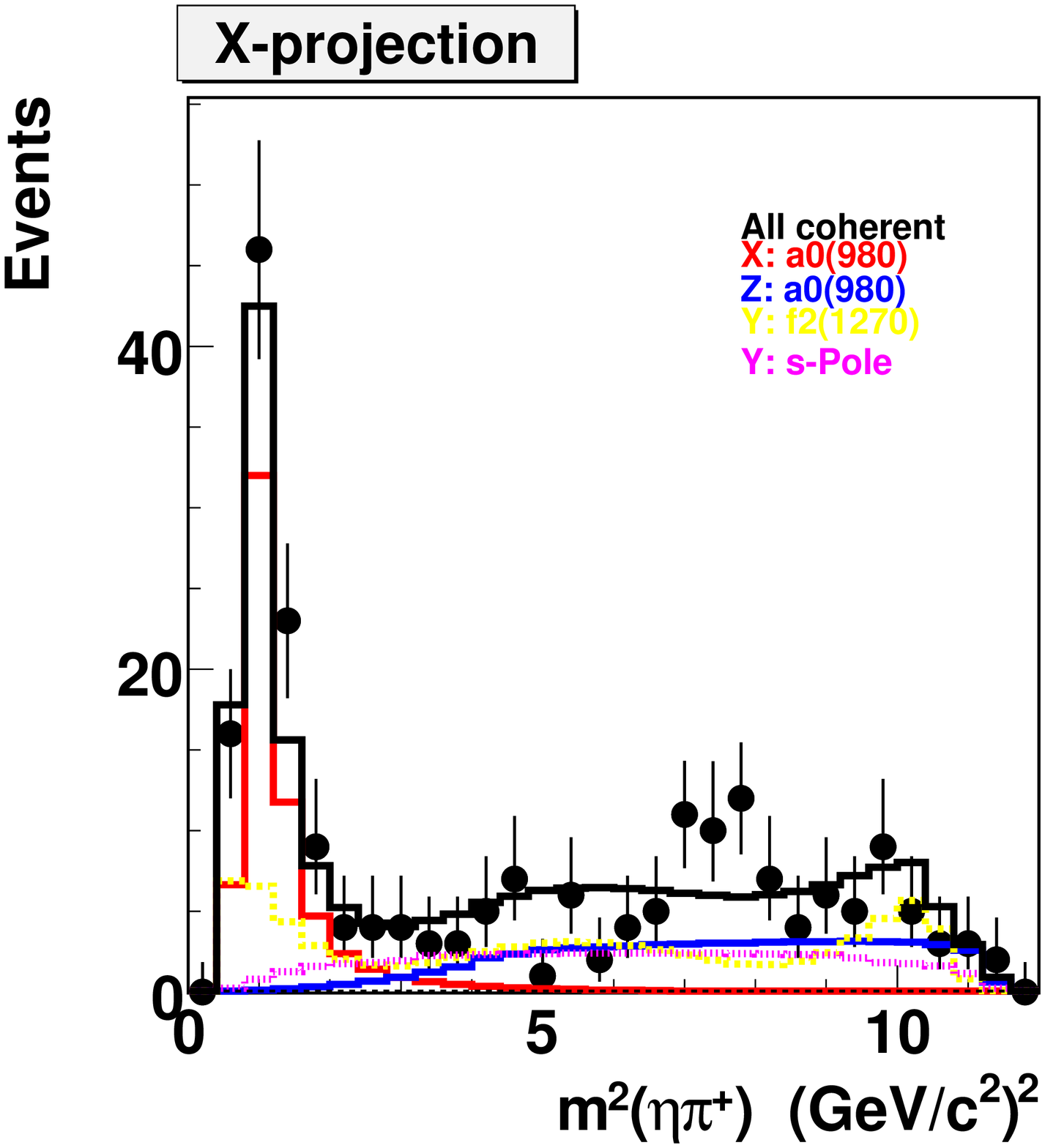} &
\includegraphics*[width=3.0in]{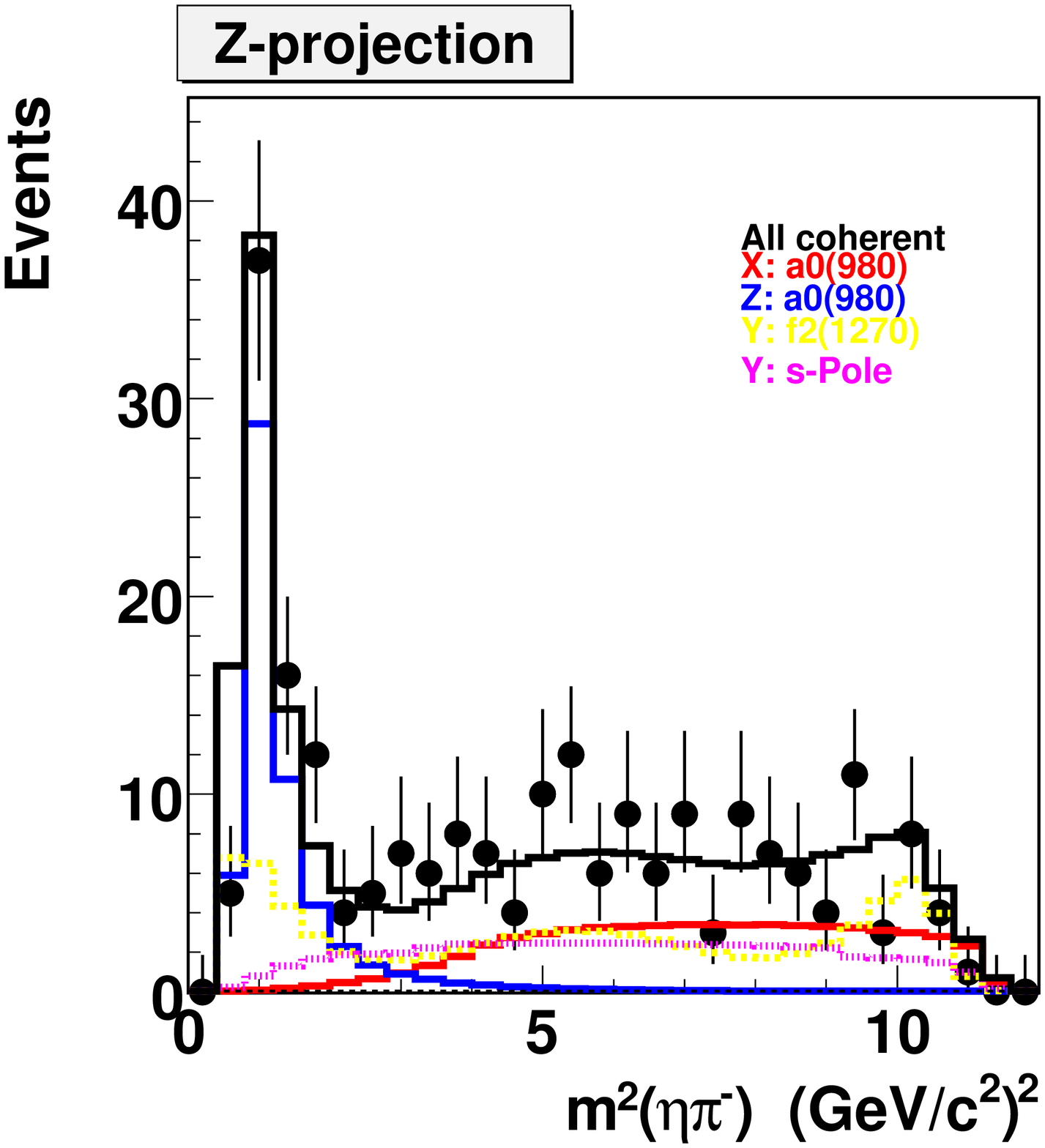} \\
\end{tabular}
\caption{Dalitz plot and projections on the three mass squared combinations
         for $\chi_{c1} \to \pi^+\pi^-\eta$.  The displayed fit projections
         are described in the text.}
\label{fig:pipietadalitz}
\end{figure}

	Figure~\ref{fig:pipietadalitz} shows the Dalitz plot
and three projections for $\chi_{c1} \to \pi^+\pi^-\eta$.
There are clear contributions from $a_0(980)^\pm \pi^\mp$ and $f_2(1270) \eta$ intermediate states,
and significant accumulation at low $\pi^+\pi^-$ mass.
Note that the $a_0(980)$ can contribute in two decay modes to the Dalitz plot.  
An isospin Clebsch-Gordan decomposition for this decay,
given in Appendix~\ref{sec:Clebsch-Gordan}, Equation~\ref{eqn:case_C_intermediate}, shows that 
amplitudes and strong phases of both charge-conjugated states should be equal.  
The overall amplitude normalization and one phase 
are arbitrary parameters and we set
$a_{a_0(980)^+} = a_{a_0(980)^-} = 1$,
$\phi_{a_0(980)^+} = \phi_{a_0(980)^-} =0$.
All other fit components are defined with respect
to these choices for $a_0(980)$. 

Our initial fit to this mode includes only $a_0(980)^\pm \pi^\mp$ and $f_2(1270) \eta$
contributions, but has a low probability of describing the data, 0.13\%, 
due to the accumulation of events at low $\pi^+\pi^-$ mass.
To account for this we try $K^0_S$, $f_0(980)$, $\rho(770)$, and $\sigma$ resonances.
Only the $\rho(770)$ and the $\sigma$ give high fit probability, 49\% and 58\% respectively. 
However the decay $\chi_{c1} \to \rho(770) \eta$ is C-forbidden,
and the low mass distribution is not well represented by $\rho(770)$ which
only gives an acceptable fit due to its large width and 
the limited statistics of our sample. 
The $\sigma$ describes well the low $\pi^+\pi^-$ mass spectrum,
and we describe the Dalitz plot with 
$a_0(980)^\pm \pi^\mp$, $f_2(1270) \eta$, and $\sigma \eta$ contributions.
Table~\ref{tab:pipietadalitz} gives the preliminary results of this fit which has a 
probability to match the data of 58.1\%.
\begin{table}
\caption{Preliminary fit results for $\chi_{c1} \to \eta\pi^+\pi^-$ Dalitz plot analysis.  The uncertainties
         are statistical and systematic.}
\begin{tabular}{l|l|l|c}
Contribution           & Amplitude               & Phase (${}^\circ$) & Fit Fraction (\%)  \\ \hline
$a_0(980)^\pm \pi^\mp$ & 1                       & 0                  & $56.2\pm3.6\pm1.4$ \\
$f_2(1270) \eta$       & $0.186\pm0.017\pm0.003$ & $-118\pm10\pm4$    & $35.1\pm2.9\pm1.8$ \\    
$\sigma \eta$          & $0.68\pm0.07\pm0.05$    & $-85\pm18\pm15$    & $21.7\pm3.3\pm0.5$ 
\end{tabular}
\label{tab:pipietadalitz} 
\end{table}
Variations to this nominal fit give the systematic uncertainties shown in the table.
We allow the 2D-efficiency to vary with its polynomial coefficients
constrained by the results of the fit to the simulated
events; the mass of the $a_0(980)$ and its coupling constants are allowed to float,
the parameters of the $\sigma$-pole are allowed to float, and we allow 
additional contributions
from $\rho(770) \eta$, $f_0(980) \eta$, $K^0_S \eta$, and $\pi_1(1400) \eta$ .  The deviation
from the nominal fit over these variants gives the systematic uncertainties shown
in Table~\ref{tab:pipietadalitz}.
For the additional contributions we do
not observe amplitudes that are significant and we limit their fit fractions
at the 95\% confidence level
to $\rho(770) \eta < 12$\%, $f_0(980) \eta < 3.5$\%, $K^0_S \eta < 0.8$\%,
$\pi_1(1400) \eta < 3.2$\%.  We note that with higher statistics this mode
may offer one of the best measurements of the parameters of the $a_0(980)$.

\begin{figure}
\begin{tabular}{cc}
\includegraphics*[width=3.0in]{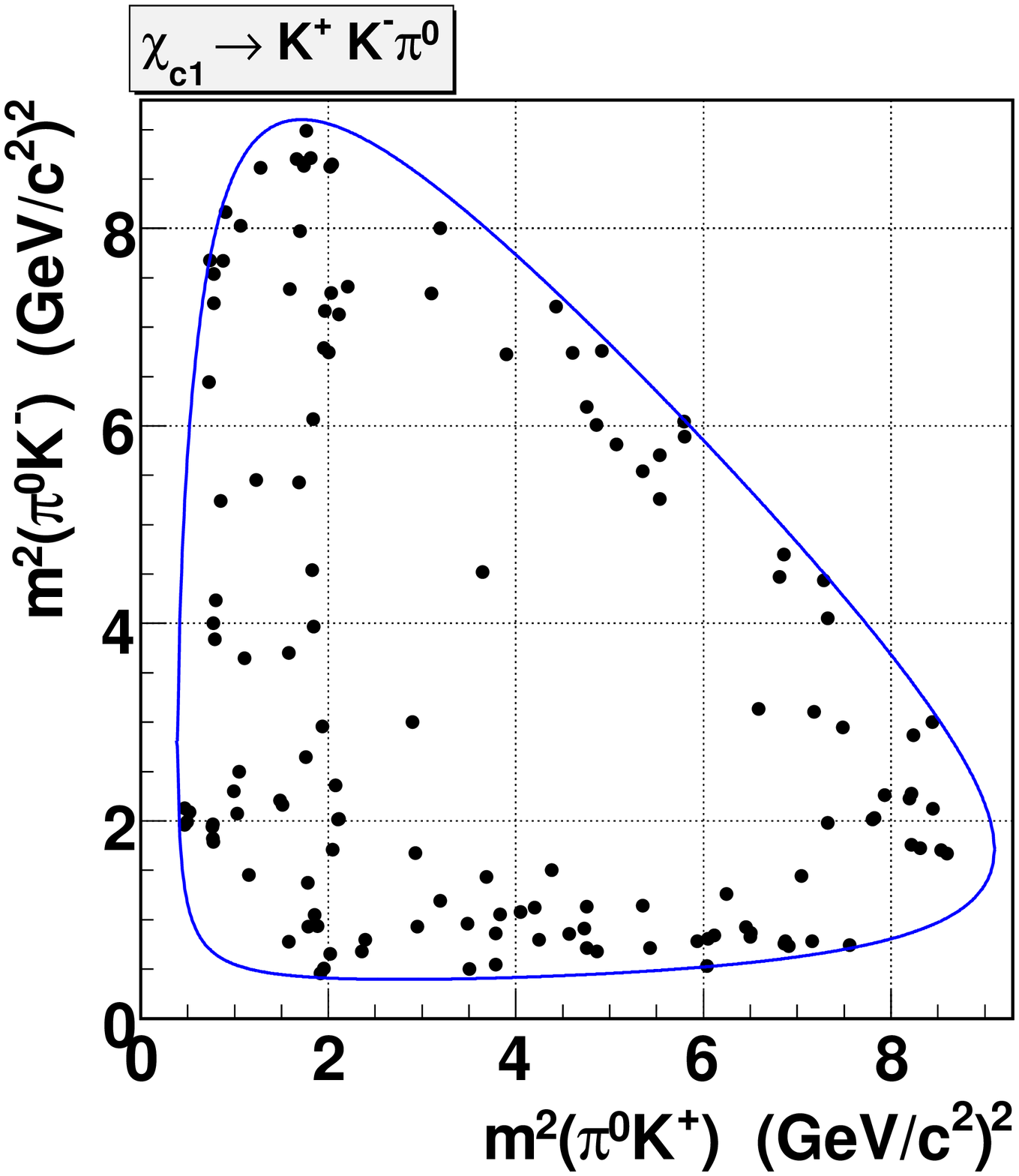} &
\includegraphics*[width=3.0in]{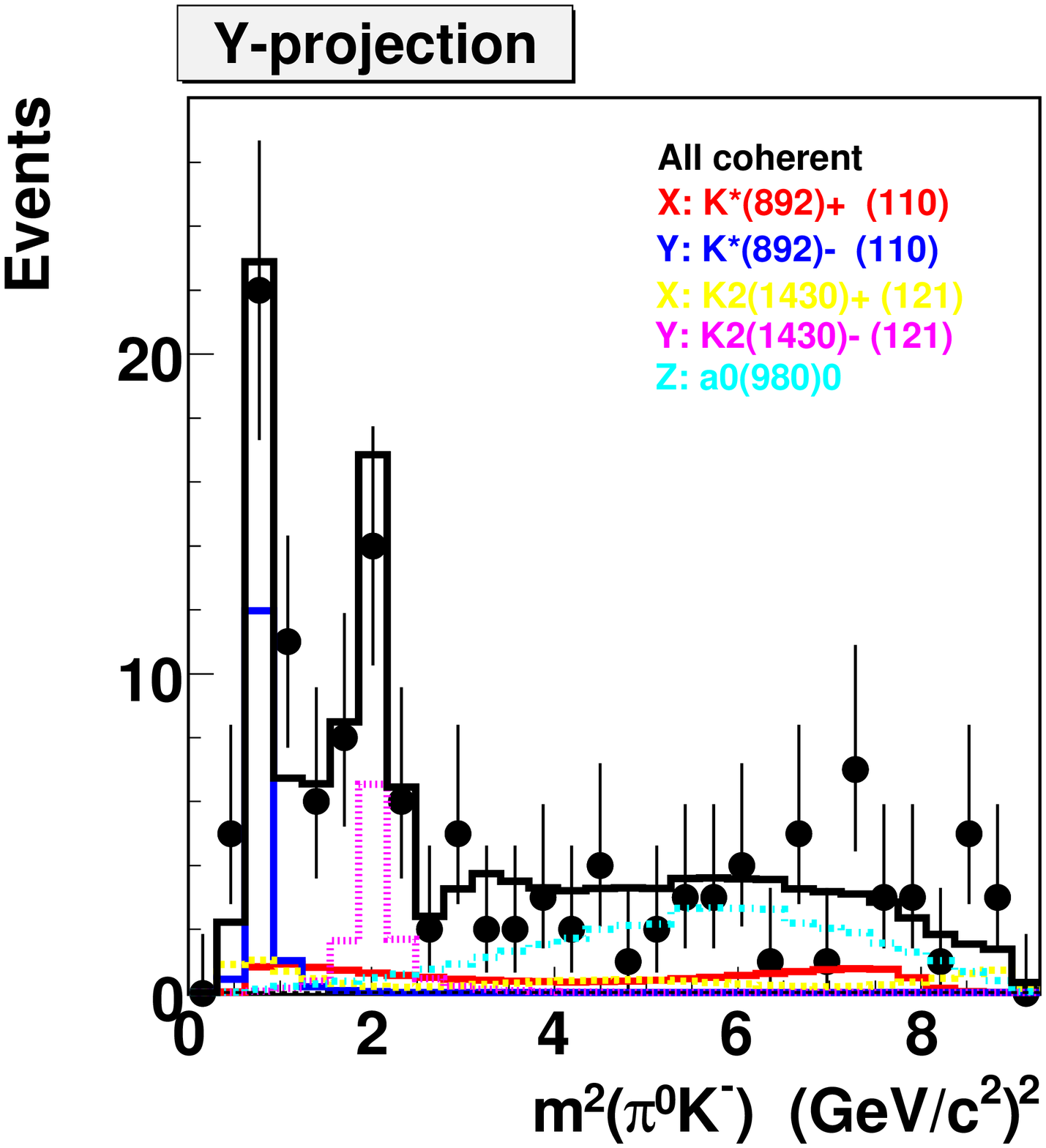} \\
\includegraphics*[width=3.0in]{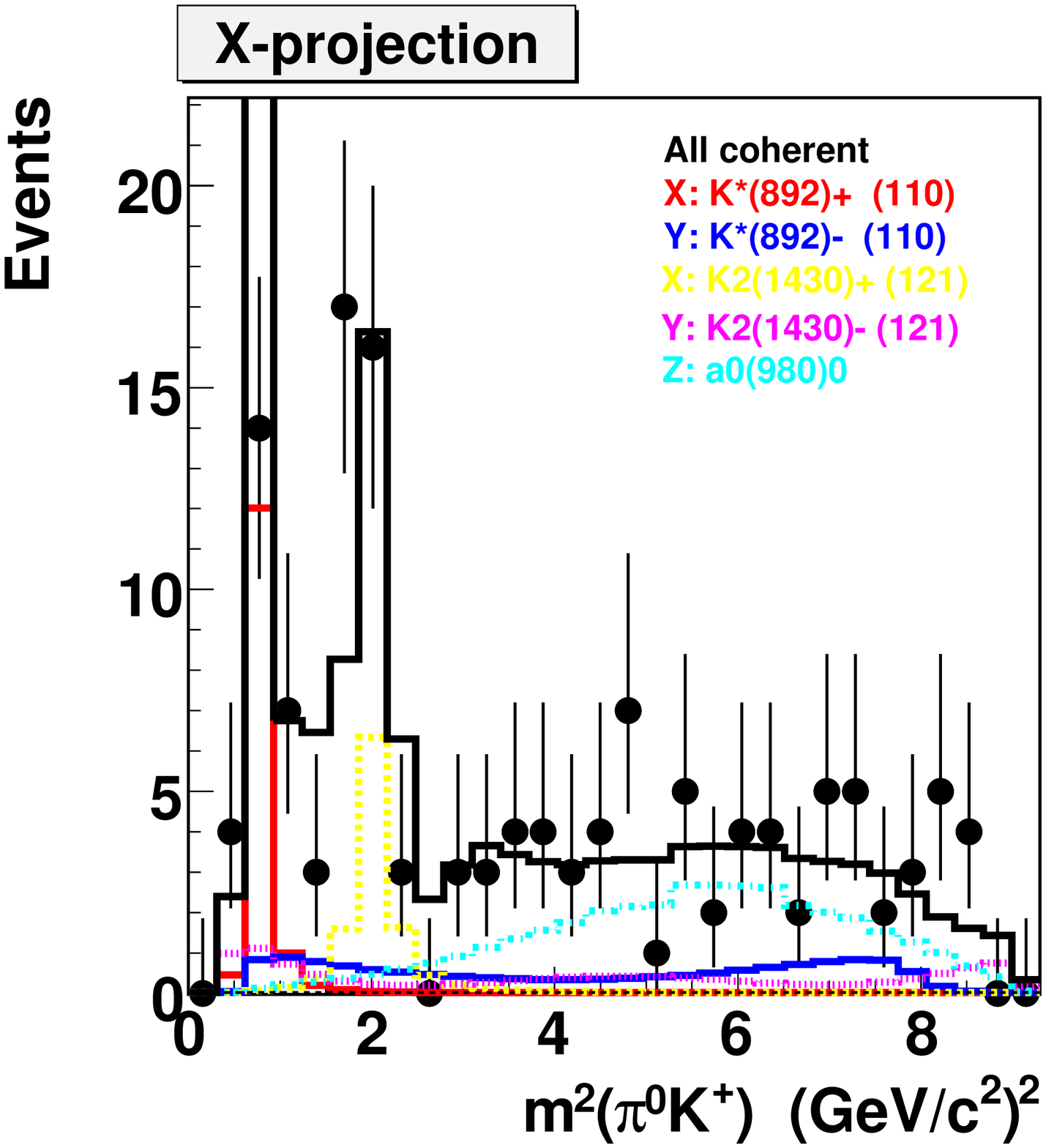} &
\includegraphics*[width=3.0in]{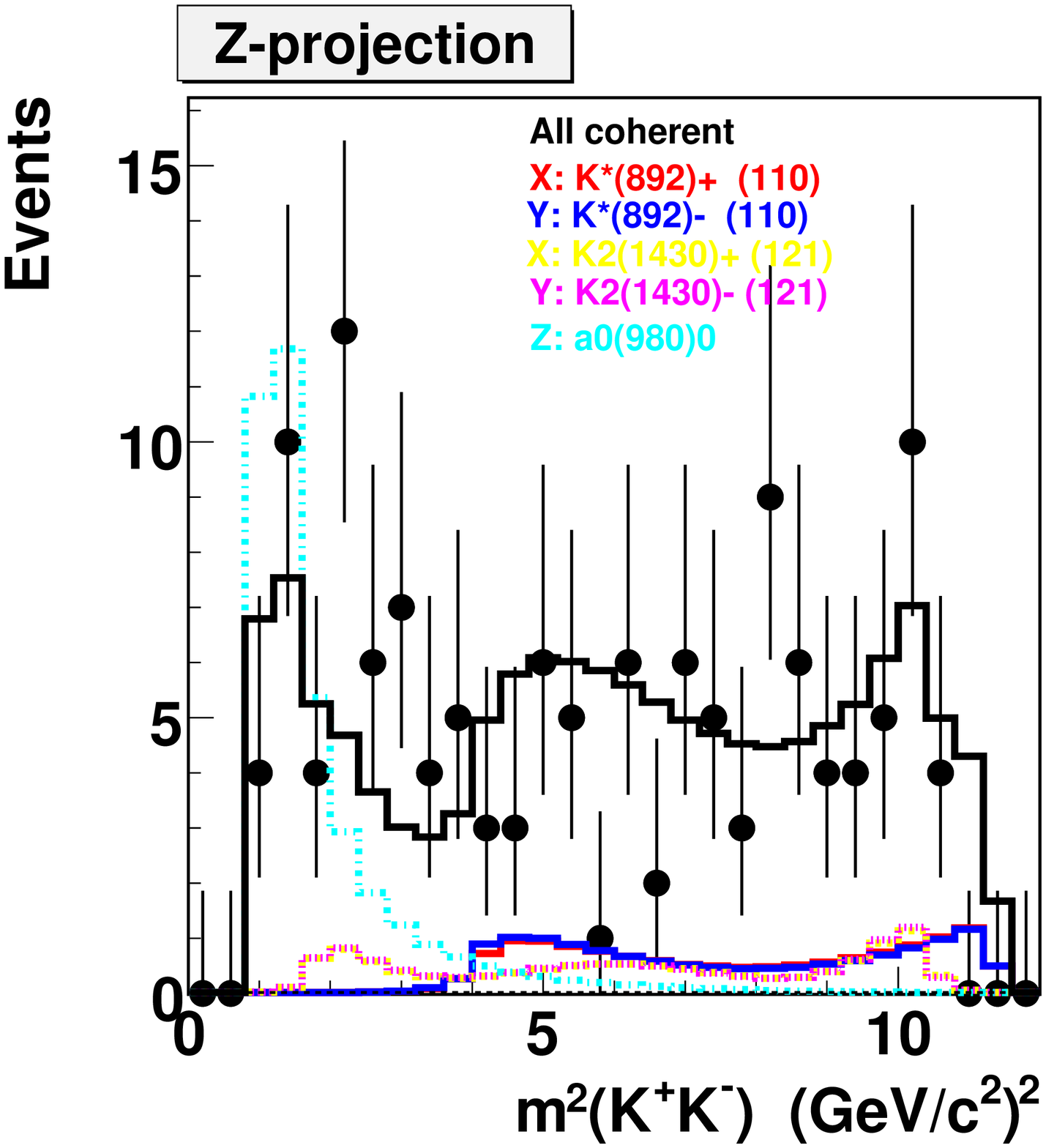} \\
\end{tabular}
\caption{Dalitz plot and projections on the three mass squared combinations
         for $\chi_{c1} \to K^+K^-\pi^0$.  The displayed fit projections
         are described in the text.}
\label{fig:KKpi0dalitz}
\end{figure}

\begin{figure}
\begin{tabular}{cc}
\includegraphics*[width=3.0in]{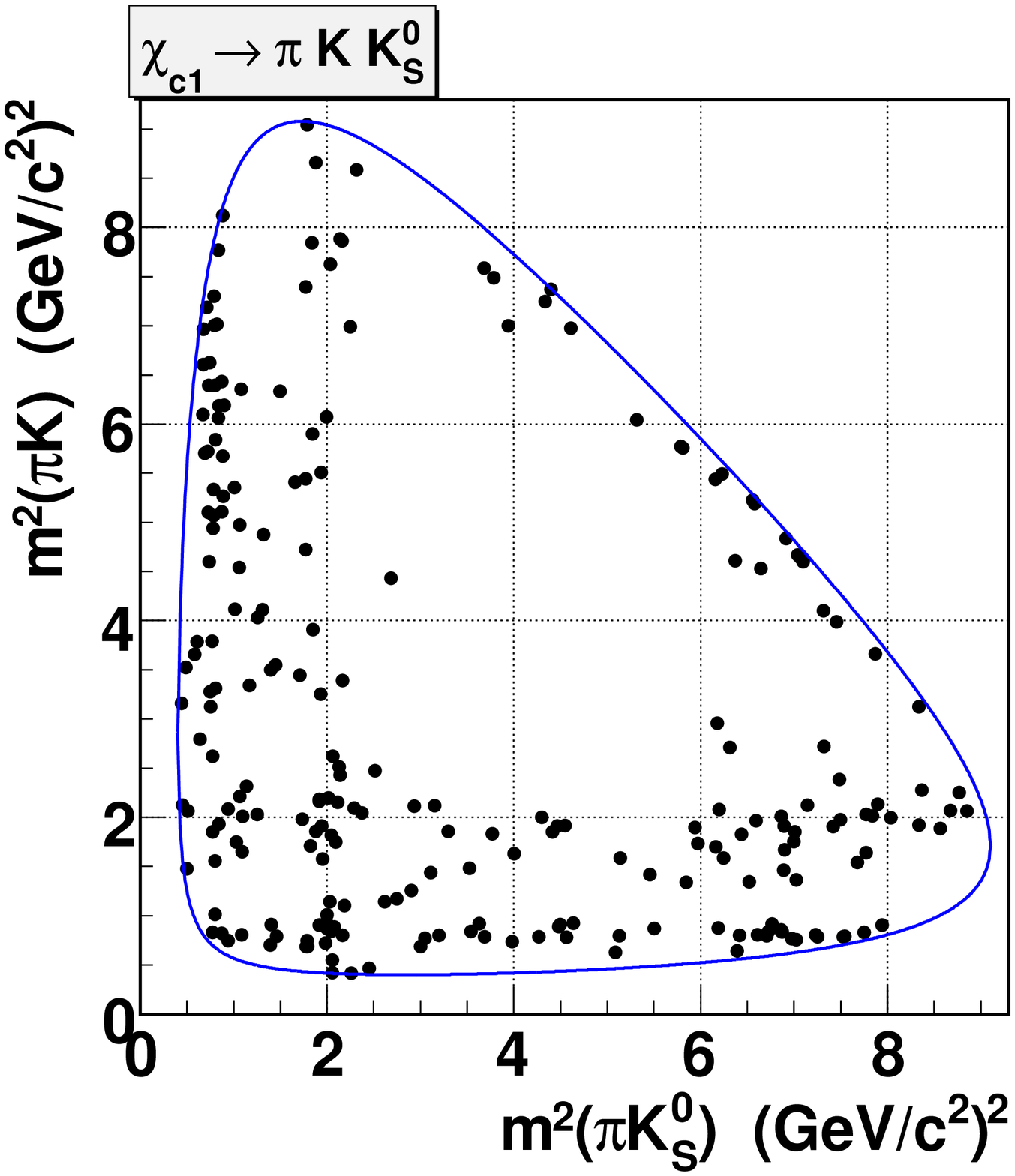} &
\includegraphics*[width=3.0in]{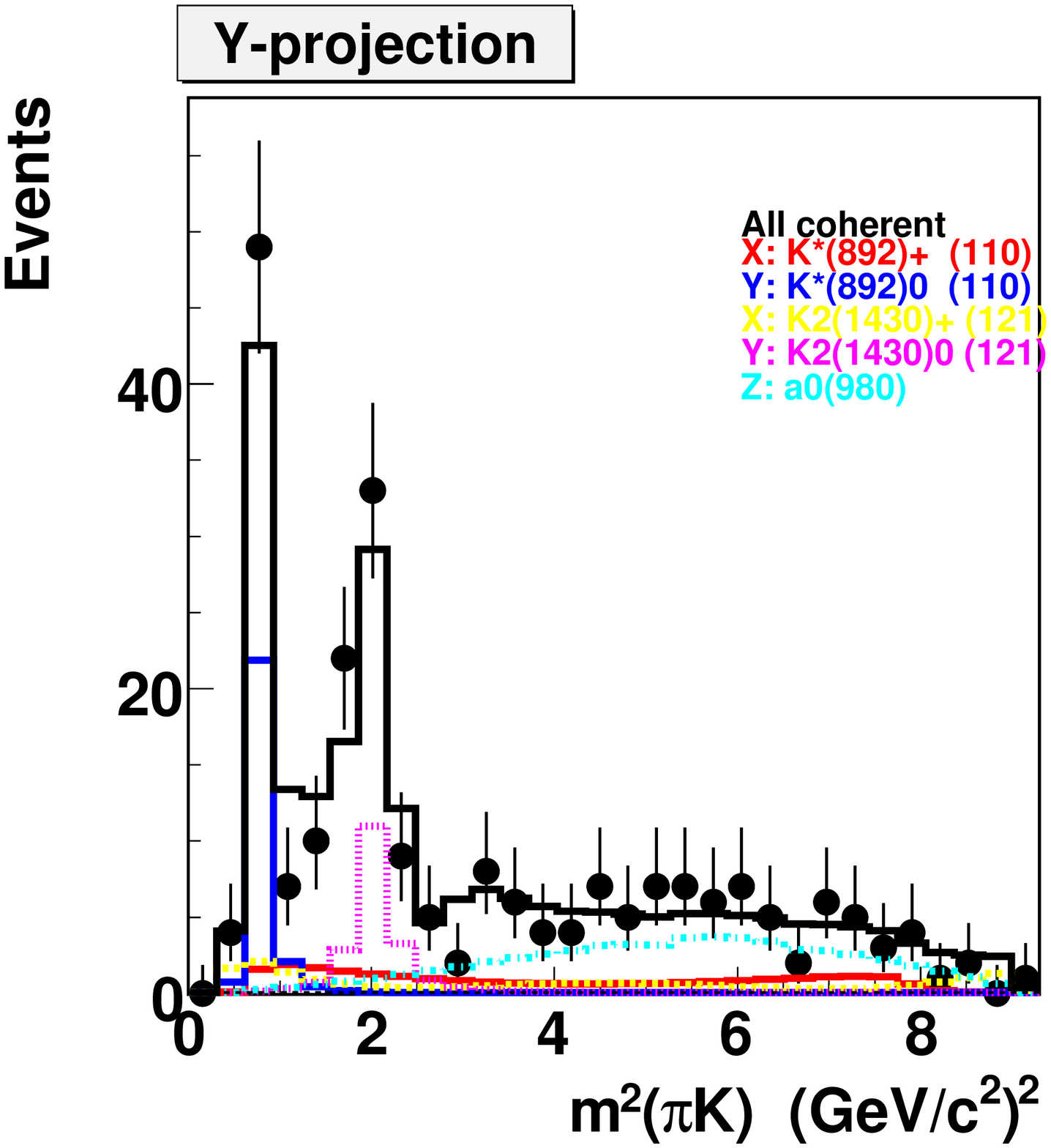} \\
\includegraphics*[width=3.0in]{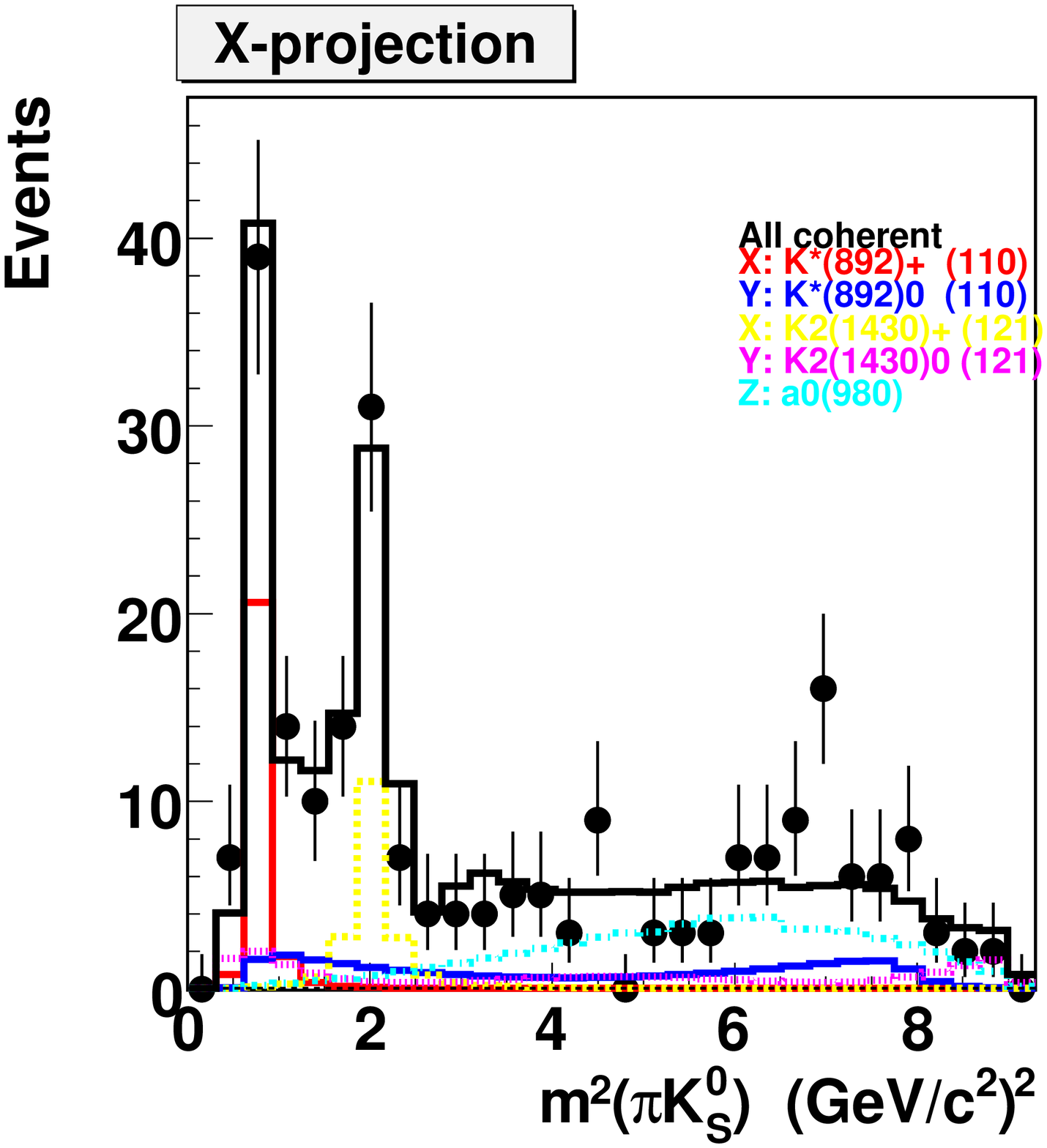} &
\includegraphics*[width=3.0in]{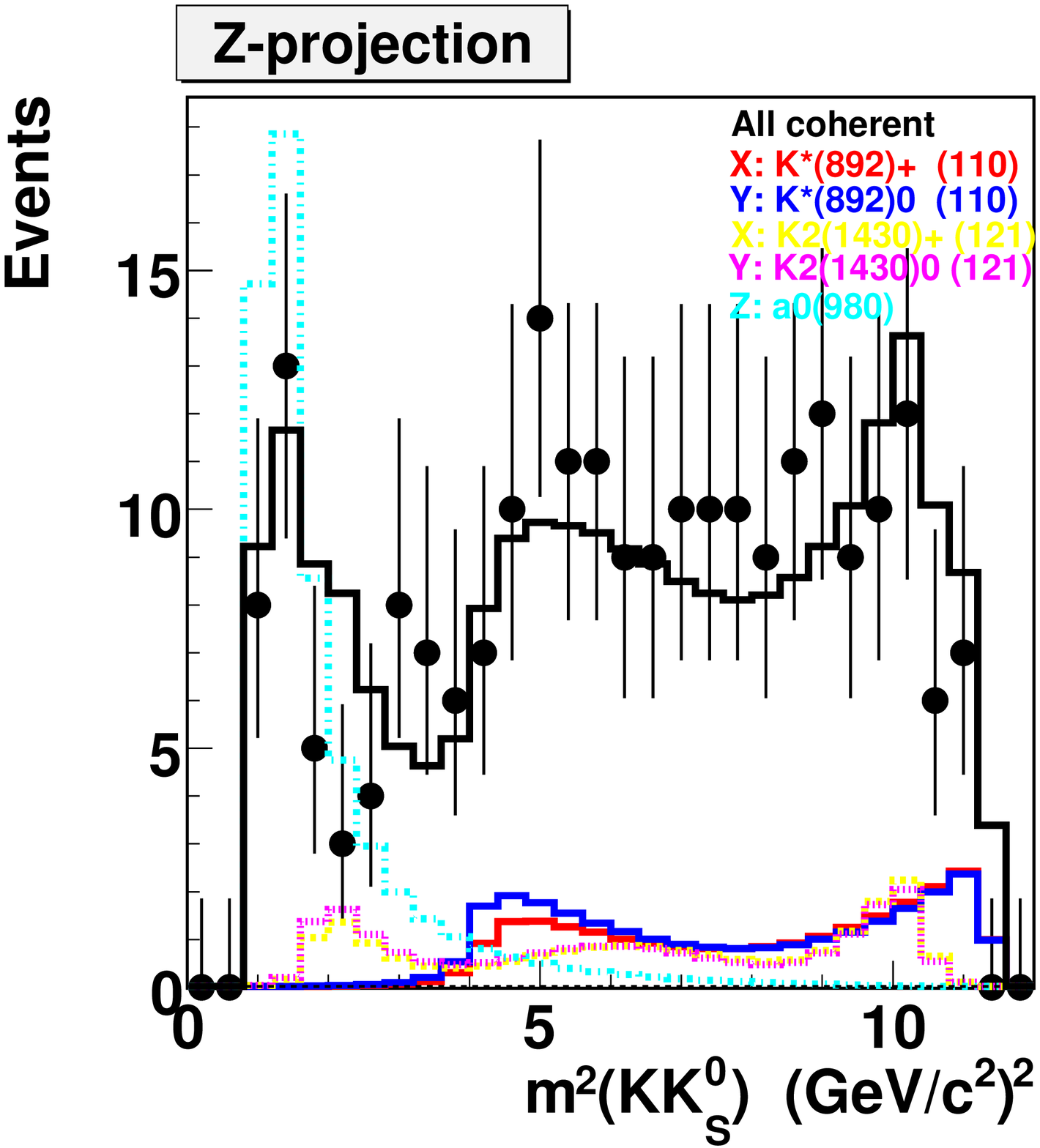} \\
\end{tabular}
\caption{Dalitz plot and projections on the three mass squared combinations
         for $\chi_{c1} \to \pi^+K^-K^0_S$.  The displayed fit projections
         are described in the text.}
\label{fig:piKK0dalitz}
\end{figure}

The Dalitz plot for $\chi_{c1} \to K^+K^-\pi^0$ decay and
its projections are shown in Figure~\ref{fig:KKpi0dalitz}, and
for $\chi_{c1} \to \pi^+K^-K^0_S$ in Figure~\ref{fig:piKK0dalitz}.
We do a combined Dalitz plot analysis to these modes taking advantage
of isospin symmetry.  An isospin Clebsch-Gordan decomposition for these decays,
Appendix~\ref{sec:Clebsch-Gordan}, Equations~\ref{eqn:case_A_FS},~\ref{eqn:case_B_FS}, 
and~\ref{eqn:case_C}, shows that these two Dalitz plots should have 
the same set of amplitude and phase parameters for all  
$K^*\overline{K}$ and $a_0(980)\pi$ intermediate states. 
The relative factor $-\sqrt{2}$
between two Dalitz plot amplitudes does not matter due to the 
individual normalization of their probability density functions.
In the combined fit to these two Dalitz plots, 
we use the following constraints on amplitudes and phases:
$a_{K^{*+}} = a_{K^{*-}} = a_{K^{*0}} = a_{\overline{K^{*0}}} \equiv a_{K^{*}}$,
$\phi_{K^{*+}} = \phi_{K^{*-}} = \phi_{K^{*0}} = \phi_{\overline{K^{*0}}} \equiv \phi_{K^{*}}$,
$a_{a(980)^+} = a_{a(980)^-} = a_{a(980)^0} \equiv a_{a(980)}$, and
$\phi_{a(980)^+} = \phi_{a(980)^-} = \phi_{a(980)^0} \equiv \phi_{a(980)}$.
The overall amplitude normalization and one phase are arbitrary parameters and
we set
$a_{K^*(892)} = 1$ and
$\phi_{K^*(892)} = 0$.

	The limited size of this sample, even in the combined Dalitz plot analysis,
and the many possible contributing resonances
leave us unable to draw clear conclusions.  Visual inspection shows apparent
contributions from $K^*(892)^\pm K^\mp$, $K^*(892)^0K^0_S$,
$K^*(1430)^\pm K^\mp$, $K^*(1430)^0K^0_S$, 
$a_0(980)^0\pi^0$, and $a_0(980)^\pm\pi^\mp$.
It is not clear if the $K^*(1430)$ are $K^*_0$ or $K^*_2$, and many
other $K\pi$ and $KK$ resonances can possibly contribute.  Our best fit
preliminary result is shown in Table~\ref{tab:KKpiDalitz} showing
\begin{table}[!htb]
\caption{Preliminary results of the combined fits to the 
$\chi_{c1} \to K^+ K^- \pi^0$ and $\chi_{c1} \to K^0_S K \pi$ Dalitz plots.
}
\begin{tabular}{l|l|l|c}
Contribution           & Amplitude               & Phase (${}^\circ$) & Fit Fraction (\%)   \\ \hline
$K^*(892)K$            & 1                       & 0                  & $19.7\pm4.0\pm2.0$  \\
$K^*_2(1430)K$         & $0.50\pm0.09\pm0.12$    & $-2\pm13\pm6$      & $18.0\pm6.6\pm3.2$  \\
$K^*_0(1430)K$         & $5.3\pm1.0\pm0.1$       & $77\pm12\pm16$     & $36.0\pm12.8\pm3.0$ \\
$K^*(1680)K$           & $2.3\pm0.5\pm0.5$       & $-38\pm12\pm12$    & $11.2\pm5.4\pm2.7$  \\
$a_0(980)\pi$          & $10.8\pm1.2\pm1.2$      & $-112\pm12\pm3$    & $29.5\pm7.1\pm2.6$
\end{tabular}
\label{tab:KKpiDalitz}
\end{table}
%
statistical and systematic errors.  
This fit has a good probability of matching the data, 32.1\%,
and agrees with fits done to the separate Dalitz plots not taking advantage of
isospin symmetry.  Unfortunately we can change our decay model by swapping the
$K^*(1430)$ plus $K^*(1680)$ contributions for a flat non-resonant background,
or a $\kappa$ and get fits of lower, but still acceptable, probability
of matching the data.  We also find that an alternative solution using the
same set of contributions fits the data acceptably, but less well than the 
displayed result.  This alternative result disagrees with
the nominal result by more than the statistical uncertainties should allow.
We conclude that our $\chi_{c1} \to KK\pi$ 
sample is too small to extract the resonant substructure.
We do clearly see contributions
from $K^*(892)K$ and $a_0(980)\pi$ at roughly the 20\% and 30\% level respectively.
The balance of the Dalitz plots are probably from higher mass $K\pi$
resonances.

	In conclusion we have searched for and studied selected three body hadronic decays
of the $\chi_{c0}$, $\chi_{c1}$, $\chi_{c2}$ produced in radiative decays
of the $\psi(2S)$ in $e^+e^-$ collisions observed with the CLEO detector.  
Our preliminary observations and branching fraction limits are summarized in Table~\ref{tab:Branching_fractions}.
In $\chi_{c1} \to \pi^+\pi^-\eta$ we have studied the resonant substructure using
a Dalitz plot analysis, and our preliminary results are summarized in Table~\ref{tab:pipietadalitz}.
Similarly in $\chi_{c1} \to KK\pi$ we clearly see contributions
from $K^*(892)K$ and $a_0(980)\pi$ at roughly the 20\% and 30\% level respectively.

We gratefully acknowledge the effort of the CESR staff
in providing us with excellent luminosity and running conditions.
D.~Cronin-Hennessy and A.~Ryd thank the A.P.~Sloan Foundation.
This work was supported by the National Science Foundation,
the U.S. Department of Energy, and
the Natural Sciences and Engineering Research Council of Canada.


\section{Appendix: Clebsch-Gordan decomposition for $\chi_{c1}$}
\label{sec:Clebsch-Gordan}

In order to constrain amplitudes and phases in $\chi_{c1}$ decays
we use a Clebsch-Gordan decomposition
of the $\chi_{c1}$ (the state with $|I=0,I_Z=0\rangle$) for possible 
isospin subsystems:
\begin{equation} \label{eqn:Kpi}
    \chi_{c1} \to (K \pi)_{I=1/2} {\overline K},
\end{equation}
\begin{equation} \label{eqn:Kbarpi}
    \chi_{c1} \to ({\overline K} \pi)_{I=1/2} K,
\end{equation}
\begin{equation} \label{eqn:KKbar}
    \chi_{c1} \to (K {\overline K})_{I=1} \pi.
\end{equation}
Below we use Clebsch-Gordan decomposition rules, $|J,M\rangle = \sum_f c_f |m_1,m_2\rangle_f$
from Ref.~\cite{pdg}. 

\subsection{Clebsch-Gordan decomposition for $K^* \to K\pi$ decays}

We assume that $K^*$s with I=1/2 form two isodoublets: 
($K^{*+}$, $K^{*0}$) and ($\overline{K^{*0}}$, $K^{*-}$) 
with ($I_Z=\frac{1}{2}$,$I_Z=-\frac{1}{2}$) respectively.
The Clebsch-Gordan decomposition rules for isospin states 
of product particles $\mathbf{ 1 \times \frac{1}{2} }$ are: 

$|\frac{1}{2}, \frac{1}{2}\rangle = \sqrt{\frac{2}{3}}|1,-\frac{1}{2}\rangle 
                            - \sqrt{\frac{1}{3}}| 0,\frac{1}{2}\rangle$

$|\frac{1}{2},-\frac{1}{2}\rangle = \sqrt{\frac{1}{3}}|0,-\frac{1}{2}\rangle
                            - \sqrt{\frac{2}{3}}|-1,\frac{1}{2}\rangle$

\begin{equation} \label{eqn:Kstar+}
     K^{*+} = |\frac{1}{2}, \frac{1}{2}\rangle = \sqrt{\frac{2}{3}} \pi^+K^0 
                                         - \sqrt{\frac{1}{3}} \pi^0 K^+,
\end{equation}
\begin{equation} \label{eqn:Kstar0}
     K^{*0} = |\frac{1}{2},-\frac{1}{2}\rangle = \sqrt{\frac{1}{3}} \pi^0K^0 
                                         - \sqrt{\frac{2}{3}} \pi^- K^+,
\end{equation}
\begin{equation} \label{eqn:Kstar0bar}
     \overline{K^{*0}}
            = |\frac{1}{2}, \frac{1}{2}\rangle = \sqrt{\frac{2}{3}} \pi^+K^- 
                                         - \sqrt{\frac{1}{3}} \pi^0 \overline{K^0},
\end{equation}
\begin{equation} \label{eqn:Kstar-}
     K^{*-} = |\frac{1}{2},-\frac{1}{2}\rangle = \sqrt{\frac{1}{3}} \pi^0K^- 
                                         - \sqrt{\frac{2}{3}} \pi^- \overline{K^0}.
\end{equation}

\subsection{Cases of $\chi_{c1} \to (K \pi)_{I=1/2} {\overline K}$ and  
                     $\chi_{c1} \to ({\overline K} \pi)_{I=1/2} K$  decays }
For $\chi_{c1} \to K^*\overline{K}$ and 
    $\chi_{c1} \to \overline{K^*} K$ modes we use 
$\mathbf{ \frac{1}{2} \times \frac{1}{2} }$ rule: 
$|0,0\rangle=\frac{1}{\sqrt{2}}(|\frac{1}{2},-\frac{1}{2}\rangle - |-\frac{1}{2},\frac{1}{2}\rangle )$
\begin{equation} \label{eqn:case_A}
     \chi_{c1} \to \frac{1}{\sqrt{2}}(K^{*+} K^- - K^{*0} \overline{K^0}),
\end{equation}
\begin{equation} \label{eqn:case_B}
     \chi_{c1} \to \frac{1}{\sqrt{2}}(\overline{K^{*0}}K^0 - K^{*-} K^+).
\end{equation}


Combining Equations~\ref{eqn:case_A},\ref{eqn:case_B} 
with Equations~\ref{eqn:Kstar+}-\ref{eqn:Kstar-} we get
\begin{equation} \label{eqn:case_A_FS}
     \chi_{c1}\sqrt{2} \to 
               \sqrt{\frac{2}{3}} \bigg[ (\pi^+K^0)K^- + (\pi^-K^+)\overline{K^0} \bigg]
             - \sqrt{\frac{1}{3}} \bigg[ (\pi^0K^+)K^- + (\pi^0K^0)\overline{K^0} \bigg],
\end{equation}
\begin{equation} \label{eqn:case_B_FS}
     \chi_{c1}\sqrt{2} \to 
               \sqrt{\frac{2}{3}} \bigg[ (\pi^+K^-)K^0 + (\pi^-\overline{K^0})K^+ \bigg]
             - \sqrt{\frac{1}{3}} \bigg[ (\pi^0K^-)K^+ + (\pi^0\overline{K^0})K^0 \bigg].
\end{equation}

Assuming charge symmetry the amplitudes in 
Equations~\ref{eqn:case_A_FS} and \ref{eqn:case_B_FS} should be equal.
From these equations we may get the ratio of rates:
\begin{equation} \label{eqn:case_rate_R1}
    \Gamma(\chi_{c1}\to \pi^+K^-K^0) /
    \Gamma(\chi_{c1}\to \pi^0K^+K^-) = 2, 
\end{equation}
\begin{equation} \label{eqn:case_rate_R2}
    \Gamma(\chi_{c1}\to \pi^-K^+\overline{K^0}) /
    \Gamma(\chi_{c1}\to \pi^0K^+K^-) = 2,
\end{equation}
or their sum
\begin{equation} \label{eqn:case_width}
       \Gamma(\chi_{c1}\to \pi^+K^-K^0)
   +   \Gamma(\chi_{c1}\to \pi^-K^+\overline{K^0}) 
   = 4 \Gamma(\chi_{c1}\to \pi^0K^+K^-).
\end{equation}

\subsection{Case of $\chi_{c1} \to (K {\overline K})_{I=1} \pi$ decay}
For $\chi_{c1} \to a\pi$ modes we use the $\mathbf{ 1\times 1 }$ rule: 
$|0,0\rangle=\frac{1}{\sqrt{3}}(|1,-1\rangle - |0,0\rangle + |-1,1\rangle )$
\begin{equation} \label{eqn:case_C_intermediate}
     \chi_{c1} \to \frac{1}{\sqrt{3}}(a^+\pi^- - a^0\pi^0 + a^-\pi^+)
\end{equation}
For $a\to K{\overline K}$ we use the $\mathbf{\frac{1}{2} \times \frac{1}{2}}$ rules:

$a^+$: $|1,1\rangle=|\frac{1}{2},\frac{1}{2}\rangle$
\begin{equation} \label{eqn:case_a+}
         a^+ \to K^+ \overline{K^0},
\end{equation}

$a^0$: $|1,0\rangle=\frac{1}{\sqrt{2}}(|\frac{1}{2},-\frac{1}{2}\rangle + |-\frac{1}{2},\frac{1}{2}\rangle )$
\begin{equation} \label{eqn:case_a0}
         a^0 \to \frac{1}{\sqrt{2}} (K^+ K^- + K^0 \overline{K^0}),
\end{equation}

$a^-$: $|1,-1\rangle=|-\frac{1}{2},-\frac{1}{2}\rangle$
\begin{equation} \label{eqn:case_a-}
         a^- \to K^0 K^-.
\end{equation}
Combining Equation~\ref{eqn:case_C_intermediate} 
with Equations~\ref{eqn:case_a+}-\ref{eqn:case_a-} we get

\begin{equation} \label{eqn:case_C}
     \chi_{c1}\sqrt{3} \to 
                      (K^+ \overline{K^0})\pi^- 
                    - \frac{1}{\sqrt{2}} \bigg[(K^+ K^-)\pi^0  + (K^0 \overline{K^0})\pi^0 \bigg]
                    + (K^0 K^-)\pi^+.
\end{equation}

From Equations~\ref{eqn:case_C} we may get the ratio
of rates:
\begin{equation} \label{eqn:case_rate_R1_for_a0}
    \Gamma(\chi_{c1}\to \pi^+K^-K^0) /
    \Gamma(\chi_{c1}\to \pi^0K^+K^-) = 2, 
\end{equation}
\begin{equation} \label{eqn:case_rate_R2_for_a0}
    \Gamma(\chi_{c1}\to \pi^-K^+\overline{K^0}) /
    \Gamma(\chi_{c1}\to \pi^0K^+K^-) = 2,
\end{equation}
or their sum
\begin{equation} \label{eqn:case_width_for_a0}
       \Gamma(\chi_{c1}\to \pi^+K^-K^0)
   +   \Gamma(\chi_{c1}\to \pi^-K^+\overline{K^0}) 
   = 4 \Gamma(\chi_{c1}\to \pi^0K^+K^-).
\end{equation}

\subsection{Consequences for Dalitz plot analysis}

Comparing 
Equations~\ref{eqn:case_rate_R1}-\ref{eqn:case_width}
for intermediate states with $K^*$ 
and
Equations~\ref{eqn:case_rate_R1_for_a0}-\ref{eqn:case_width_for_a0}
for intermediate states with $a(980)$,
one may notice that they are identical.
Thus observations of $\pi^+K^-K^0$ and 
charge conjugated $\pi^-K^+\overline{K^0}$ final states 
on the same Dalitz plot will yield the same ratio
between all $K^*$ and $a(980)$ amplitudes 
for both $K^0K^-\pi^+$ and $K^+K^-\pi^0$ Dalitz plots. 
The ratio of amplitudes between these two Dalitz plots is  $-\sqrt{2}$.
The relative negative sign between amplitudes
does not matter, because the rates depend on the matrix element squared.
The relative factor $\sqrt{2}$ between amplitudes 
does not matter, because the normalizations are different for
each Dalitz plot.
This isospin analysis implies that these two Dalitz plots, 
$K^0K^-\pi^+$ and $K^+K^-\pi^0$,
can be parametrized using a common set of parameters for each 
of $K^*$s and $a(980)$ intermediate state with
\begin{equation}
a_{K^{*+}} = a_{K^{*-}} = a_{K^{*0}} = a_{\overline{K^{*0}}} \equiv a_{K^{*}} ,
\end{equation}
\begin{equation}
\phi_{K^{*+}} = \phi_{K^{*-}} = \phi_{K^{*0}} = \phi_{\overline{K^{*0}}} \equiv \phi_{K^{*}},
\end{equation}
\begin{equation}
a_{a(980)^+} = a_{a(980)^-} = a_{a(980)^0} \equiv a_{a(980)},  
\end{equation}
\begin{equation}
\phi_{a(980)^+} = \phi_{a(980)^-} = \phi_{a(980)^0} \equiv \phi_{a(980)}.
\end{equation}


\section{Appendix: The effect of $\chi_{c1}$ polarization}
\label{sec:angular_distributions}
\subsection{Angular distributions} 
In this analysis we use the angular distributions explicitly shown in 
Table~\ref{tab:angular_distributions}
derived for a {\it non-polarized} decaying particle ($\chi_{c1}$)
in Ref.~\cite{Filippini-Fontana-Rotondi}.
\begin{table}[!htb]
\caption{\label{tab:angular_distributions} Angular distributions.}
\begin{center}
\begin{tabular}{c|c}
\hline
$J \to j+L$ & Probability \\
\hline
0$\to$0+0 & uniform \\
1$\to$0+1 & uniform \\
2$\to$0+2 & uniform \\
0$\to$1+1 & $(1+z^2)\cos^2\theta$ \\
0$\to$2+2 & $(z^2+3/2)^2(\cos^2\theta-1/3)^2$ \\
1$\to$1+0 & $1 + z^2\cos^2\theta$ \\
1$\to$1+1 & $\sin^2\theta$ \\
1$\to$1+2 & $1 + (3 + 4 z^2)\cos^2\theta$ \\
1$\to$2+1 & $(1+z^2) \big[1 + 3\cos^2\theta + 9 z^2 (\cos^2\theta - 1/3)^2 \big]$ \\
1$\to$2+2 & $(1 + z^2)\cos^2\theta \sin^2\theta$     (*)\\
2$\to$1+1 & $3 + ( 1 + 4 z^2)\cos^2\theta$ \\
2$\to$1+2 & $\sin^2\theta$ \\
2$\to$2+0 & $1 + z^2 / 3 + z^2 \cos^2\theta + z^4 (\cos^2\theta - 1/3)^2$ \\
2$\to$2+1 & $1 + z^2 / 9 + (z^2/3 - 1)\cos^2\theta - z^2 (\cos^2\theta - 1/3)^2$ \\
2$\to$2+2 & $1 + z^2 / 9 + (z^2/3 - 1)\cos^2\theta + (16z^4+21z^2+9)(\cos^2\theta - 1/3)^2 / 3$ (*)\\
\hline
\end{tabular}
\end{center}
(*) These formulas have been derived based on the covariant helicity formalism approach
    also discussed in Ref.~\cite{Filippini-Fontana-Rotondi}.
\end{table}
The parameters in Table~\ref{tab:angular_distributions}
follow the conventions of the original publication:
          $J$ - spin of the initial particle  $J \to j + c$;
          $j$ - spin of the resonance $j \to a + b$, where $a,b,c$ - spin 0 particles;
          $L$ - orbital angular momentum between ($j^p$) resonance and the recoil particle $c$.
A relativistic correction factor $z$ and $\cos\theta$, 
where $\theta$ is an angle between directions of particles $a$ and $c$ in resonance $j^p$ rest frame
is discussed in Appendix~\ref{sec:z_and_cosTheta}.

In order to account for the polarization of the initial $\chi_{c1}$  
produced in the radiative decay $\psi(2S) \to \gamma \chi_{c1}$, a 
full Partial Wave Analysis formalism, for example from Ref.~\cite{BES_PWA_formalism},
would be needed.
The matrix element amplitudes would depend on additional angular variables
in addition to the two invariant squared masses as in the regular Dalitz plot analysis.
The number of parameters would also be increased to account for 
different helicity amplitudes contributions.
On the other hand, an expected $\chi_{c1}$ production angular distribution, 
$\propto 1 - \frac{1}{3} \cos^2\theta_{\chi_{c1}}$, is $\sim$15\% different
from uniform and we expect that the effect of polarization is small
in the decays under study.

\subsection{Expressions for relativistically non-invariant variables 
in terms of Dalitz plot invariant variables}
\label{sec:z_and_cosTheta}

Although a covariant spin tensor formalism is applied, for simplicity the formulas in
Ref.~\cite{Filippini-Fontana-Rotondi} and
Table~\ref{tab:angular_distributions}
are expressed in terms of relativistically 
non-invariant variables $z$ and $\cos\theta$. 
Here we discuss the meaning of these variables and their
expression in terms of particle masses and invariant masses.

Assuming the decay $d\to Rc$, $R\to ab$ we may derive 
kinematic variables used in formulas for angular distribution 
in terms of Dalitz plot invariant variables.
The energy and momentum of particle $a$ in the resonance $R$ or $(ab)$ rest frame
is a trivial expression
\begin{equation} \label{eqn:E_a}
        E_a = \frac{m_{ab}^2 + m_a^2 - m_b^2}{2m_{ab}}, ~~~~ P_a=\sqrt{E_a^2-m_a^2}.
\end{equation}
The energy and momentum of particle $c$ in the resonance $R$ rest frame
can be obtained from the invariant mass $m_d^2$ expression in the $R$ rest frame.
Indeed, $m_d^2 = (p_c + p_R)^2$,  
in the resonance $R$ rest frame $p_c=(E_c,\vec{P_c})$, $p_R = (m_R,\vec{0})$, 
$\to m_d^2 = m_c^2 + m_R^2 + 2 m_R E_c$,
hence
\begin{equation} \label{eqn:E_c}
        E_c = \frac{m_d^2 - m_{ab}^2 - m_c^2}{2m_{ab}}, ~~~~ P_c=\sqrt{E_c^2-m_c^2}.
\end{equation}
Now, the $\cos\theta$ between directions of particles $a$ and $c$ in resonance $R$ rest frame
can be expressed through the known energies and momenta of particles $a$ and $c$ 
and their measured invariant mass squared $m_{ac}^2$:
\begin{equation} \label{eqn:cosTheta}
        \cos\theta = \frac{m_a^2 + m_c^2 + 2 E_a E_c - m_{ac}^2}{2 P_a P_c}.
\end{equation}
The resonance energy and relativistic correction factor 
(used in in Table~\ref{tab:angular_distributions})
in the decaying particle $d$ rest frame are:
\begin{equation} \label{eqn:E_r_and_Z}
          E_r = \frac{m_d^2 + m_{ab}^2 - m_c^2}{2 m_d}, ~~~~
          z^2 = \gamma^2_R - 1 = \frac{E_r^2}{m_{ab}^2} - 1.
\end{equation}


\begin{thebibliography}{99}
\bibitem{cleog} We use a GEANT-3 based full detector response Monte Carlo simulation.
\end{thebibliography}
\end{document}